\documentclass[journal]{IEEEtran}
\IEEEoverridecommandlockouts
\ifCLASSINFOpdf
\else
   \usepackage{graphicx}
   \graphicspath{{../eps/}}
   \DeclareGraphicsExtensions{.eps}
\fi
\usepackage{amsthm,amsmath,amssymb}
\usepackage{mathrsfs}
\usepackage{amsfonts}
\usepackage{graphicx, amssymb}
\usepackage{amsthm}
\usepackage{amsmath}
\usepackage{epsfig}
\usepackage{graphicx}
\usepackage{graphics}
\usepackage{multirow}
\usepackage{threeparttable}
\usepackage{amsthm}
\usepackage{float}
\usepackage{color}
\usepackage{algorithm}
\usepackage{algpseudocode}
\usepackage{bm}
\usepackage{algorithmicx}
\usepackage{algpseudocode}
\usepackage{setspace}
\usepackage{authblk}
\usepackage{cite}
\usepackage{pifont}
\usepackage{xcolor}
\usepackage{epstopdf}
\usepackage{caption}
\usepackage{graphicx}
\usepackage{float}
\usepackage{subcaption}
\usepackage{cleveref}
\theoremstyle{definition}

\newcommand{\RNum}[1]{\uppercase\expandafter{\romannumeral #1\relax}}

\usepackage{wrapfig}
\usepackage{psfrag}
\usepackage{graphicx}
\usepackage{threeparttable}
\usepackage{cases}
\usepackage{subeqnarray}
\usepackage{color}
\usepackage{underscore}

\usepackage{textcomp}
\usepackage{xcolor}
\def\BibTeX{{\rm B\kern-.05em{\sc i\kern-.025em b}\kern-.08em
    T\kern-.1667em\lower.7ex\hbox{E}\kern-.125emX}}

\begin{document}
\title{Efficient Channel Estimation for Millimeter Wave and Terahertz Systems Enabled by Integrated Super-resolution Sensing and Communication}
\author{
Jingran Xu, Huizhi Wang, Yong Zeng, \IEEEmembership{Senior Member,~IEEE}, Xiaoli Xu,
Qingqing Wu, \IEEEmembership{Senior Member,~IEEE}, Fei Yang, Yan Chen, \IEEEmembership{Senior Member,~IEEE}, and Abbas Jamalipour, \IEEEmembership{Fellow,~IEEE}~

\thanks{This work was supported by the National Natural Science Foundation of China with grant number 62071114.
Part of this work has been presented at the 2024 IEEE Wireless Communications and Networking Conference (WCNC) Workshops, Dubai, United Arab Emirates, 21-24 Apr. 2024 \cite{conference}.}
\thanks{Jingran Xu, Huizhi Wang, Yong Zeng, and Xiaoli Xu are with the National Mobile Communications Research
Laboratory and Frontiers Science Center for Mobile Information Communication and Security, Southeast University, Nanjing 210096, China. Yong Zeng is also
with the Purple Mountain Laboratories, Nanjing 211111, China (e-mail:
{jingranxu,~wanghuizhi,~yong_zeng, xiaolixu}@seu.edu.cn). (\emph{Corresponding author: Yong Zeng.})

Qingqing Wu is with Department of Electronic Engineering, Shanghai Jiao
Tong University, Shanghai 200240, China (e-mail: qingqingwu@sjtu.edu.cn).

Fei Yang and Yan Chen are with the Wireless Technology Lab., 2012 Lab, Shanghai Huawei Technologies Co., Ltd., Shanghai 201206, China (e-mail: {yangfei8, bigbird.chenyan}@huawei.com).

Abbas Jamalipour is with the School of Electrical and Computer Engineering, The University of Sydney, Camperdown, NSW 2006, Australia (e-mail: a.jamalipour@ieee.org).}
}

\maketitle

\begin{abstract}
Integrated super-resolution sensing and communication (ISSAC)
has emerged as a promising technology to achieve extremely high precision sensing for those key parameters, such as the angles of the sensing targets.
In this paper, we propose an efficient channel estimation scheme enabled by ISSAC for
millimeter wave (mmWave) and TeraHertz (THz) systems with a hybrid analog/digital beamforming architecture,
where both the pilot overhead and the cost of radio frequency (RF) chains are significantly reduced.
The key idea is to exploit the fact that subspace-based super-resolution algorithms such as multiple signal classification (MUSIC) can estimate channel parameters accurately without requiring dedicate a priori known pilots.
In particular, the proposed method consists of two stages. First, the angles of the multi-path channel components are estimated in a pilot-free manner during the transmission of data symbols.
Second, the multi-path channel coefficients are estimated with very few pilots.
Compared to conventional channel estimation schemes that rely solely on channel training, our approach requires the estimation of much fewer parameters in the second stage.
Furthermore, with channel multi-path angles obtained, the beamforming gain can be achieved when pilots are sent to estimate the channel path gains. To comprehensively investigate the performance of the proposed scheme, we consider both the basic line-of-sight (LoS) channels and more general multi-path channels.
We compare the performance of the minimum mean square error (MMSE) of channel estimation
and the resulting beamforming gains of our proposed scheme with the traditional scheme that rely exclusively on channel training.
It is demonstrated that our proposed method significantly outperforms the benchmarking scheme.
Simulation results are presented to validate our theoretical findings.
\end{abstract}

\begin{IEEEkeywords}
Channel estimation, hybrid beamforming, integrated super-resolution sensing and
communication (ISSAC), little pilot
\end{IEEEkeywords}

\IEEEpeerreviewmaketitle
\section{Introduction}

Integrated sensing and communication (ISAC) has been identified as one of the six usage scenarios for
IMT-2030 (6G) \cite{IMT2030}.
By applying super-resolution algorithms in ISAC systems, the concept of integrated super-resolution sensing and communication (ISSAC) is further developed \cite{zhangchaoyue}, which achieves exceptionally high sensing performance for key parameters such as angle, delay, and Doppler of sensing targets.
In addition, due to the spatial sparsity caused by severe
propagation loss, high-frequency channels such as millimeter wave (mmWave) and TeraHertz (THz) are generally dominated by a limited number of multi-path components. Consequently, those channels can be
modeled parametrically, focusing on the path angles of departure/arrival (AoD/AoA) and the corresponding path gains \cite{mmwavechannel1}. Therefore,
high-frequency channel estimation can be converted into the estimation of multi-path angles and corresponding channel gains, without the need to directly estimate the high-dimensional channel matrix.

Massive multiple input multiple output (MIMO) has drawn tremendous attention
in mmWave and THz systems, which requires massive radio frequency (RF) chains with extremely
high power consumption and hardware cost if conventional fully digital signal processing architectures are used \cite{RFchain1,RFchain2,RFchain3}.
To reduce the cost of RF chains attached to antennas, the hybrid analog/digital beamforming structure has been extensively studied for antenna architecture design \cite{hybrid1,hybrid2,hybrid3,hybrid4}.
In this beamforming structure, since the receiver cannot
access the signals of all antenna elements, the corresponding spatial covariance matrix of all antennas cannot be directly obtained.
As a consequence, the classical multiple signal classification (MUSIC) algorithm in ISSAC systems cannot be directly applied.
In addition, traditional estimation of signal parameters via rotational invariance technique (ESPRIT) and spatial smoothing technique are no longer applicable due to the destruction of the shift-invariance of array response. Beamspace MUSIC which can be employed in hybrid structure was studied in \cite{beamspaceMUSIC} to find directions of coherent sources. However, the spatial smoothing therein is still performed in the digital structure.
To achieve AoA estimation, \cite{power1,power2,power3} investigated a method by utilizing the maximum received power.
However, due to the restriction of Rayleigh limitation \cite{Rayleigh},
these approaches are unable to obtain super-resolution AoA estimation.
To use MUSIC algorithm, \cite{BSA} proposed a beam sweeping algorithm (BSA) to
reconstruct the spatial covariance matrix by solving linear equations.

To fully exploit the potential gain provided by the massive MIMO systems,
it is critical to obtain accurate channel state information (CSI).
For conventional pilot-based channel estimation schemes \cite{estimationmethod},
least squares (LS) and linear minimum mean square error (LMMSE) schemes can be straightforwardly executed
by examining the correlation between the known pilot sequences and received signals.
Maximum likelihood estimation (MLE) provides superior performance but is more complex by employing a likelihood-based method to maximize the probability of observed signals based on the known pilots.
However, the increasing number of antennas in massive MIMO significantly raises the dimension of the channel matrix,
which poses a severe challenge to traditional channel estimation due to the increased length of pilot sequence
and computational complexity.
To address the challenge of CSI acquisition, channel knowledge map (CKM) is proposed
to enable a promising paradigm shift from the
traditional environment-unaware communications to the novel
environment-aware communications \cite{CKMtutorial,CKMmagazine}.
In addition, for the estimation of mmWave and THz channels,
a straightforward and low-complexity approach is to search in the angular space by varying the steering directions of the beamformer \cite{qiongsou1,qiongsou2}, since mmWave and THz channels consist of only a few dominant components. However, the exhaustive search
leads to high training overhead.
To alleviate the high pilot overhead,
the compressive sensing (CS) algorithms have been proposed, exploiting the angular sparsity of massive MIMO channels \cite{CS1,CS2,xujie}.
However, it is non-trivial to find a suitable dictionary matrix and
the low signal-to-noise ratio (SNR) and imperfect measurement feedback from users to
base station (BS) result in additional performance loss.
On the other hand, to avoid directly dealing with the channel matrix, the
methods of channel parameter estimation is also effective in angular channel models
by using the classical spatial spectrum estimation (SSE) algorithms such as discrete Fourier
transformation (DFT) \cite{gaofeifei1,gaofeifei2}, MUSIC algorithm \cite{esmusic1,esmusic2} and ESPRIT \cite{esESPRIT1,esESPRIT2}.

However, current channel parameter estimation methods based on the SSE algorithms
still rely heavily on transmitting pilot signals to estimate angles \cite{gaofeifei1,gaofeifei2,esmusic1,esmusic2,esESPRIT1,esESPRIT2},
which requires significant pilot overhead.
In this paper, for mmWave and THz communications employing hybrid beamforming,
we propose a highly efficient channel estimation scheme enabled by ISSAC, which is able to achieve
accurate CSI estimation with very few pilots.
The critical idea is that subspace-based super-resolution algorithms, such as MUSIC, can achieve accurate
estimation of channel parameters by transmitting information-bearing symbols rather than dedicated pilots.
Specifically, without using any pilots, the angles of the multi-path channel components
are first estimated during the phase of data transmission.
Subsequently, by utilizing the hybrid beamforming gain based on the obtained angles,
the multi-path channel coefficients can be further estimated
with very few pilots since much fewer parameters
need to be estimated compared to traditional channel estimation methods.
Our main contributions are summarized as follows:
\begin{itemize}
	\item First, we model an uplink mmWave and THz communication system using a hybrid analog/digital
beamforming structure. For the conventional channel estimation method, to obtain the upper bound of signal
detection performance, we study the scenario of fully digital beamforming and analyze the performance
for channel estimation and signal detection. It is revealed that high SNR and estimation accuracy can only be guaranteed when the pilot sequence is sufficiently long, which incurs
high overhead and reduces the spectral efficiency.

\item Next, to significantly reduce the pilot overhead while
still guaranteing the communication performance, we propose an efficient channel estimation scheme enabled by ISSAC.
For the proposed method, no dedicated pilot is needed to estimate the angles of the multi-path channel components,
and only very few pilots are required for the complex-valued path coefficients.

\item  Furthermore, we analyze the performance of the proposed ISSAC-enabled channel
estimation scheme.
The MMSE of channel estimation and the resulting beamforming gains are first analyzed for the basic line of sight (LoS) channels. The study is then extended to more general multi-path channels, where both the fully digital and hybrid analog/digital structure are studied.
It is demonstrated that the proposed method can achieve higher receive SNR and higher estimation accuracy than the traditional scheme.
The pilot overhead is further compared for achieving the same performance in terms of the expected SNR.
Numerical results are presented to validate our theoretical analysis.

\end{itemize}

The rest of this paper is organized as follows. Section II
presents the system model of mmWave and THz communication with a hybrid analog/digital beamforming structure,
and analyzes the performance of the traditional channel estimation scheme.
The proposed efficient channel estimation scheme enabled by ISSAC is further introduced.
Section III evaluates the performance of our proposed scheme by analyzing the MMSE of channel estimation and the resulting beamforming gains for both the LoS and more general multipath channel models.
In Section IV, numerical results are presented to validate our theoretical studies.
Finally, we conclude the paper in Section V.

\emph{Notations:} In this paper, italic letters denote scalars.
The boldface lower- and upper-case letters denote vectors and matrices, respectively.
$\mathbf{a}^{\mathrm{T}}$, $\mathbf{a}^{\mathrm{H}}$ and $\| \mathbf{a}\|$
give the transpose, Hermitian transpose, and Euclidean norm of a vector $\mathbf{a}$, respectively.
${{\mathbf{A}}^{*}}$, ${{\mathbf{A}}^{\mathrm{T}}}$ and ${{\mathbf{A}}^{\mathrm{H}}}$ denote the conjugate, transpose and Hermitian transpose of a matrix $\mathbf{A}$, respectively.
$\mathrm{vec}(\mathbf{A})$ represents stacking the columns of matrix $\mathbf{A}$ into a column vector.
${\rm tr}(\mathbf{B})$ denotes the trace of a square matrix $\mathbf{B}$.
$\otimes$ refers to the Kronecker product.
${\mathbf{I}_{M}}$ is an ${M\times M}$ identity matrix.
$\mathbb{C}^{M\times N}$ denotes the space of ${M\times N}$ matrices with complex entries.
$\mathbb{E}_{X}[\cdot]$ is the expectation taken over the random variable $X$.
$\mathcal{CN}(\mu,\sigma^{2})$ denotes the
distribution of a circularly symmetric complex Gaussian (CSCG) variable with mean $\mu$ and covariance matrix $\sigma^{2}$.
$o\left( x \right)$ is an infinitesimal of $x$.
\section{System Descriptions}
\subsection{System Model}
\begin{figure}[!t]
  \centering
  \centerline{\includegraphics[width=3.5in,height=1.2in]{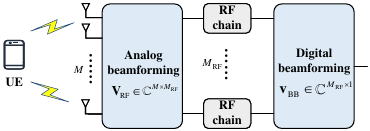}}
  \caption{A mmWave/THz communication system with hybrid analog/digital beamforming.}
  \label{SIMOsystem model}
  \end{figure}
For the traditional channel estimation scheme depending on channel training \cite{howmuchtraining},
the length of pilots is required to be at least the number of transmit antennas.
To satisfy this constraint and avoid the huge pilot overhead in the downlink system, we study an uplink mmWave and THz communication system.
As shown in Fig. \ref{SIMOsystem model}, a BS is equipped with a uniform linear array (ULA) of $M\gg 1$ antennas to serve a single-antenna user equipment (UE).
To achieve cost-effective deployment, the BS is assumed to have $M_{\mathrm{RF}}\leq M$ RF chains, where analog receive beamforming is applied by phase shifters.
The uplink receive beamforming is separated into RF
and baseband processing, denoted by $\mathbf{V}_{\mathrm{RF}}\in {{\mathbb{C}}^{M\times M_{\mathrm{RF}}}}$ and $\mathbf{v}_{\mathrm{BB}}\in {\mathbb{C}}^{M_{\mathrm{RF}}\times 1}$, respectively.
They are under the constraint of ${{\left\| {{\mathbf{V}}_{\mathrm{RF}}}{{\mathbf{v}}_{\mathrm{BB}}} \right\|}^{2}}=1$.
Amplitude and phase adjustments are both feasible for the baseband receive beamforming
$\mathbf{v}_{\mathrm{BB}}$, whereas only phase modifications can be applied to the RF receive beamforming $\mathbf{V}_{\mathrm{RF}}$ with variable phase shifters and combiners \cite{hybridbeamforming}.
Therefore, each entry of $\mathbf{V}_{\mathrm{RF}}$ is normalized to satisfy $\left| \mathbf{V}_{\mathrm{RF}}^{i,j} \right|=\frac{1}{\sqrt{M}}$,
where $\left| \mathbf{V}_{\mathrm{RF}}^{i,j} \right|$ is
the magnitude of the $(i, j)$ th element of $\mathbf{V}_{\mathrm{RF}}$.

For the classical channel estimation method based on pilot training \cite{howmuchtraining}, each channel coherence block includes two phases.
Pilots are transmitted in the first phase to estimate the channel, followed by data transmission.
Specifically, let $\rho $ and $\kappa $ respectively denote the length of pilot and data transmission sequence.
Denote by ${{\mathbf{h}}}$ the uplink channel from the UE to the BS.
$P_{t}$ and $P_{d}$ are the transmit power by the UE during pilot and data transmission phases, respectively.
Further let ${{\phi }}\left( n \right)$ denote the pilot sequence sent by
the UE satisfying $\sum\limits_{n=1}^{\rho }{{{\left| \phi  \left( n \right) \right|}^{2}}}=\rho $.
Let $s(n)$ denote the i.i.d. information-bearing symbols of the UE,
which follows CSCG distribution with normalized power, i.e.,
$s(n)\sim\mathcal{C}\mathcal{N}\left( 0, 1\right)$.
The received signals at the BS during the pilot and data symbol durations can be respectively written as
\begin{equation}
	{{{\mathbf{y}}_{t}(n)}=\sqrt{{{P}_{t}}}\mathbf{h}{{\phi }\left( n \right)}+{{\mathbf{u}}_{t}(n)},n=1,...,\rho}, \label{originalpilotphase}
\end{equation}
\begin{equation}
	{{{\mathbf{y}}_{d}(n)}=\sqrt{{{P}_{d}}}\mathbf{h}{s}\left( n \right)+{\mathbf{u}}_{d}(n)},n=\rho+1,...,\rho+\kappa, \label{originaldataphase}
\end{equation}
where ${{\mathbf{u}}_{t}(n)}$ and ${{\mathbf{u}}_{d}(n)}$ denote the i.i.d. CSCG noise with zero-mean and
variance ${{\sigma }^{2}}$.

In addition, denote the receive SNR at the BS without beamforming during the pilot and data transmission phase as ${\mathrm{SNR}}_{t}$ and ${\mathrm{SNR}}_{d}$, respectively,
i.e., ${\mathrm{SNR}}_{t}=\frac{{{P}_{t}}}{M\sigma^{2}}{{\left\| \mathbf{h} \right\|}^{2}}$,
${\mathrm{SNR}}_{d}=\frac{{{P}_{d}}}{M\sigma^{2}}{{\left\| \mathbf{h} \right\|}^{2}}$.
\subsection{Conventional Channel Estimation Method}
For traditional pilot-based channel estimation, since ${\phi }\left( n \right)$ is known at the BS, $\mathbf h$ can be estimated by performing LS method as \cite{pilotestimation}
\begin{equation}
	{{\bf{\hat h}}}_{\rm{con}}=\frac{1}{\sqrt{{{P}_{t}}{{\rho }^{2}}}}\sum\limits_{n=1}^{\rho }{{{\mathbf{y}}_{t}}\left( n \right){{\phi }^{*}}\left( n \right)}=\mathbf{h}+\frac{1}{\sqrt{{{P}_{t}}{{\rho }^{2}}}}\mathbf{u}_{t}^{\prime },\label{hLSmethod}
\end{equation}
where $\mathbf{u}_{t}^{\prime }=\sum\limits_{n=1}^{\rho }{{{\mathbf{u}}_{t}}\left( n \right)}{{\phi }^{*}}\left( n \right)$ is the resulting noise vector, which is the CSCG noise with power ${{\rho }}{{\sigma }^{2}}$, i.e., ${\mathbf{u}}_{t}^{\prime }\sim\mathcal{C}\mathcal{N}\left( \mathbf 0, {{\rho }}{{\sigma }^{2}}{\mathbf{I}}_{M}\right)$.

To gain more useful insights and obtain the upper bound of signal detection performance, we first consider the fully digital beamforming for the traditional channel estimation method.
Then, the receive beamforming vector $\mathbf{v}_{\mathrm{con}}=\mathbf{v}_{\mathrm{BB}}=\frac{{\mathbf{\hat{h}_{\mathrm{con}}}}}{\left\| {\mathbf{\hat{h}_{\mathrm{con}}}} \right\|}\in {{\mathbb{C}}^{M\times 1}}$ can be used,
where the number of RF chains ${M_{\mathrm{RF}}}=M$, and the number of data streams ${M_s}=1$.
The resulting signal can be accordingly formulated as
\begin{equation}
	{{{y}_{d,\mathrm{con}}}(n)=\mathbf{v}_{\mathrm{con}}^{\mathrm{H}}{{\mathbf{y}}_{d}(n)}
		=\sqrt{{{P}_{d}}}\mathbf{v}_{\mathrm{con}}^{\mathrm{H}}\mathbf{h}{s}\left( n \right)+u_{d,\mathrm{con}}\left( n \right)},\label{afterbeampilot}
\end{equation}
where $u_{d,\mathrm{con}}\left( n \right)=\mathbf{v}_{\mathrm{con}}^{\mathrm{H}}{{\mathbf{u}}_{d}(n)}$ denotes the CSCG noise with zero-mean and variance ${{\sigma }^{2}}$.
The expected SNR of the signal in \eqref{afterbeampilot} is
\begin{equation}
\begin{split}
	{{\gamma }_{\mathrm{con}}}& =\mathbb{E}\Big[ \frac{{{P}_{d}}{{\left| \mathbf{v}_{\mathrm{con}}^{\mathrm{H}}\mathbf{h} \right|}^{2}}}{{{\sigma }^{2}}} \Big]\\
		& =\frac{{{P}_{d}}}{{{\sigma }^{2}}}{{\mathbf{h}}^{\mathrm{H}}}\mathbb{E}\left[ {{\mathbf{v}}_{\mathrm{con}}}\mathbf{v}_{\mathrm{con}}^{\mathrm{H}} \right]\mathbf{h}.\label{gamma1initial}
\end{split}
\end{equation}

\emph{Theorem 1:}
The expected SNR in \eqref{gamma1initial} can be further approximated as
\begin{equation}
	{{{\gamma }_{\mathrm{con}}}\approx \frac{{{P}_{d}}{{\left\| \mathbf{h} \right\|}^{2}}}{{{\sigma }^{2}}}\left( 1-\chi \right)},\label{gamma1app}
\end{equation}
where $\chi=\frac{1-\frac{1}{M}}{\rho {\mathrm{SNR}}_{t}+1}$ denotes the penalty loss resulting from imperfect channel estimation.
\begin{IEEEproof}
Please refer to Appendix A.
\end{IEEEproof}
It is shown that the penalty loss $\chi$ increases monotonically with the number of antennas $M$.
This is due to the fact that the beamforming gain is not utilized in the channel training stage.
Note that the expected SNR ${{\gamma }_{\mathrm{con}}}$ in \eqref{gamma1app} increases monotonically with $\rho \mathrm{SNR}_{t}$,
and the upper bound of ${{\gamma }_{\mathrm{con}}}$ can be obtained as
\begin{equation}
{{{\gamma }_{\mathrm{con}}}<{{\gamma }_{\mathrm{upper}}}=\frac{{{P}_{d}}{{\left\| \mathbf{h} \right\|}^{2}}}{{{\sigma }^{2}}}}.\label{upper}
\end{equation}
It is reached when $\rho \mathrm{SNR}_{t}\to \infty$.


On the other hand, denote the estimation error of $\mathbf{h}$ as $\mathbf{\tilde{h}}_{\rm{con}}={{{{\bf{\hat h}}}_{{\rm{con}}}} - {\bf{h}}}$,
then the MMSE of ${\mathbf{h}}$ is
\begin{equation}
	{e_{{\rm{con}}}} = \mathbb{E}\Big[ {{{\Big\| \mathbf{\tilde{h}}_{\rm{con}}  \Big\|}^2}} \Big] = \frac{{M{\sigma ^2}}}{{{P_t}\rho }}.\label{RMSEe1}
\end{equation}
It is observed that a longer pilot length improves the estimation accuracy.

Thus, to achieve accurate channel estimation and higher receive SNR,
it is required that the pilot length is sufficiently long, which, however, will increase overhead and reduce the spectral efficiency.
As a result, we ask the following question: Is it possible to use a few number of pilots while guaranteing the communication performance?
To answer this question, the channel model in the parametric form is firstly introduced, with regard to the angles of multipath channel components and the corresponding path gains in \cite{mmwavechannel1}.

In particular, the channel from the UE to the BS can be written as $\mathbf{h}=\sum\limits_{l=1}^{L}{{{\alpha }_{l}}\mathbf{a}\left( {{\theta }_{l}} \right)}$,
where $L$ is the number of multipaths, ${{\theta }_{l}}$ is the angle of the $l$-th path,
$\mathbf{a}\left( {{\theta }_{l}} \right)$ is the BS array response vector, and ${{\alpha }_{l}}$ is the complex-valued path coefficient.
For ULA with half-wavelength spacing, we have $\mathbf{a}\left( {{\theta }_{l}} \right)={{\left[ 1,{{e}^{j\pi\sin \left( {{\theta }_{l}} \right)}},\cdots ,{{e}^{j\pi\left( M-1 \right)\sin \left( {{\theta }_{l}} \right)}} \right]}^{\mathrm{T}}}\in {{\mathbb{C}}^{M\times 1}}$.
Furthermore, by letting $\bm{\Theta}=\left({{\theta }_{1}},\cdots ,{{\theta }_{L}} \right)$, $\mathbf{A}\left(\bm{\Theta}  \right)=\left[ \mathbf{a}\left( {{\theta }_{1}} \right),\cdots ,\mathbf{a}\left( {{\theta }_{L}} \right) \right]$, and $\bm{ \alpha}={{\left[ {{\alpha }_{1}},\cdots ,{{\alpha }_{L}} \right]}^{\mathrm{T}}}$,
$\mathbf{h}$ can also be expressed as
\begin{equation}
{\mathbf{h}=\mathbf{A\left(\bm{\Theta}  \right)}\bm{ \alpha}}\label{channelh}.
\end{equation}
Therefore, the channel estimation is to estimate the angle of arrivals $\bm {\Theta}$ (AoAs) and the complex-valued path coefficient $\bm {\alpha}$.
It is worth noting that $\bm {\Theta}$ and $\bm {\alpha}$ vary with different timescales.
In particular, the complex-valued path coefficient $\bm {\alpha}$ varies between different channel coherence time,
while $\bm{\Theta}$ usually alters much slower than $\bm {\alpha}$.
It is thus assumed that $\bm{\Theta}$ remains unchanged during each path-invariant block \cite{pathinvariant1,pathinvariant3,pathinvariant4}.
\subsection{Efficient Channel Estimation Enabled by ISSAC}
In order to use a few number of pilots while guaranteeing the communication performance,
we propose an efficient channel estimation scheme enabled by ISSAC with the need of very few pilots.
Different from prior works only using dedicated pilot signals in angle-aided channel estimation \cite{gaofeifei1,gaofeifei2,esmusic1,esmusic2,esESPRIT1,esESPRIT2},
both the existing pilot and data signals are employed in our proposed method to estimate the angles of the multipath channel components by super-resolution algorithms.
In addition, since the path AoAs $\bm \Theta$ remain unaltered over several coherence time durations, the estimation result from previous coherence blocks is still applicable for the subsequent ones.
Furthermore, when estimating the channel path gains $\bm \alpha$ with pilots sent,
the beamforming gain can be enjoyed for the achieved angles during each path-invariant block, which can effectively enhance the accuracy of channel estimation.

\begin{figure}[!t]
  \centering
  \centerline{\includegraphics[width=3.6in,height=2.4in]{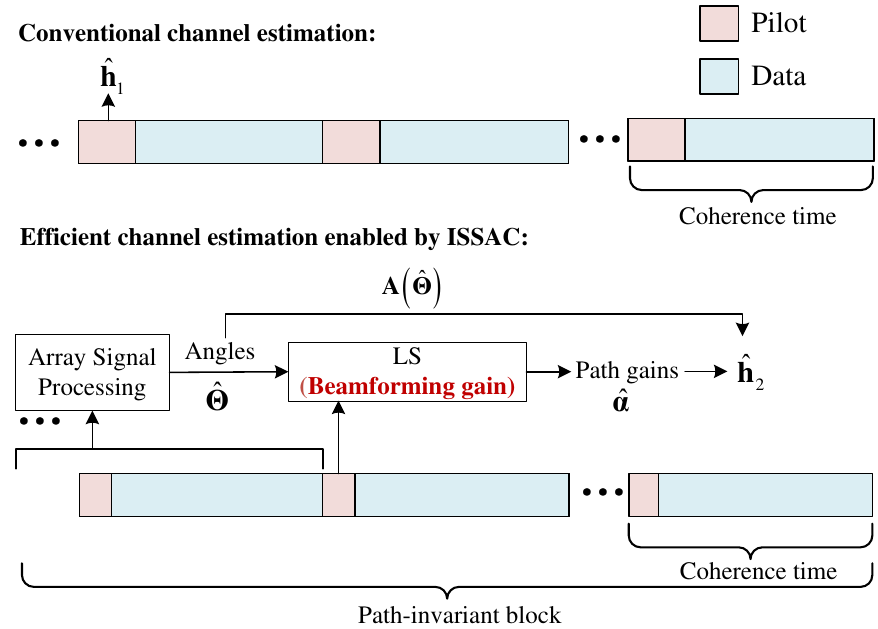}}
  \caption{Comparison of the proposed channel estimation scheme enabled by ISSAC with the traditional channel estimation scheme.}
  \label{comparison}
\end{figure}

In particular, as shown in Fig. \ref{comparison}, the proposed efficient channel estimation scheme enabled by ISSAC comprises two stages.
In the first stage, the BS estimates the directions of paths by classical sensing algorithms like the periodogram algorithm \cite{periodogram1}, spectral-based algorithms like MUSIC \cite{musicDEDI} or ESPRIT \cite{ESPRITDEDI}.
In the second stage, to estimate the multi-path channel coefficients, a few uplink pilots are sent by the UE and the receive beamforming are performed by the BS via matching the path directions estimated in previous stage.

\section{Performance Analysis of the Proposed Method}
Next we study the performance of channel estimation and signal detection for our proposed channel estimation scheme.

\subsection{Free-space LoS Channel}\label{los}
We first consider the case of LoS channel, i.e., $L=1$.
The channel vector is expressed as ${\bf h}=\alpha \mathbf{a}\left( \theta  \right)$.
We consider the codebook based analog beamforming,
where the columns of $\mathbf{V}_{\mathrm{RF}}$ are selected from the
predetermined codebooks $\mathcal{V}\in {{\mathbb{C}}^{M\times M}}$.
The total number of candidate beamforming vectors is $M$.
For simplicity, $\frac{M}{M_{\mathrm{RF}}}$ is assumed as an integer.
We further assume that the candidate beamforming vectors are chosen non-repeatedly, then the total number of possible
choices for $\mathbf{V}_{\mathrm{RF}}$ is $\frac{M}{M_{\mathrm{RF}}}$, which are denoted as $\mathbf{V}_{\mathrm{RF},i}$,
where $i=1,...,\frac{M}{M_{\mathrm{RF}}}$.

To utilize the data from several snapshots, we use the Bartlett estimation algorithm, which is a power spectra estimation scheme with the advantage of reducing the variance of the periodogram \cite{angleestimation}.
However, for hybrid digital/analog beamforming structure, the BS cannot directly
obtain the signals of all the antenna elements.
To estimate angle, we will introduce two methods to obtain spatial covariance matrix,
named digital signal reconstruction method and digital covariance matrix reconstruction method,
which are applicable to different transmission rates of data symbols.
It is noted that both methods mentioned here can also be applied to the angle estimation for multi-path channels.

On one hand, if the transmission rate of data symbols $s\left(n \right)$ is slow, which remain unchanged during $\frac{M}{M_{\mathrm{RF}}}$ analog beamforming switches,
then the digital signal reconstruction method is introduced.
The resulting signal at the BS of the $i$-th search after analog beamforming during the same data symbol is written as
\begin{equation}
\begin{split}
{{\mathbf{y}}_{\mathrm{RF},i}}(n)& =\mathbf{V}_{\mathrm{RF},i}^{\mathrm{H}}{{\mathbf{y}}_{d}}(n)\\
& =\sqrt{{{P}_{d}}}\mathbf{V}_{\mathrm{RF},i}^{\mathrm{H}}\mathbf{h}s\left(n \right)+{{\mathbf{u}}_{\mathrm{RF},i}}\left(n \right).\label{yrfd}
\end{split}
\end{equation}
By concatenating ${{\mathbf{y}}_{\mathrm{RF},i}}(n)$ in \eqref{yrfd} for all $i = 1,\cdots, \frac{M}{M_{\mathrm{RF}}}$,
we have ${{\mathbf{y}}_{\mathrm{all},\mathrm{RF}}}(n)={{\left[ \mathbf{y}_{\mathrm{RF},1}^{\mathrm{T}}(n),\cdots ,\mathbf{y}_{\mathrm{RF},\frac{M}{{{M}_{\mathrm{RF}}}}}^{\mathrm{T}}(n) \right]}^{\mathrm{T}}}\in {{\mathbb{C}}^{ M\times 1}}$.
Similarly, let $\mathcal{V}={\left[ {{\mathbf{v}}_{\mathrm{RF},1}},\cdots ,{{\mathbf{v}}_{\mathrm{RF},\frac{M}{{{M}_{\mathrm{RF}}}}}} \right]}$,
${{\mathbf{u}}_{\mathrm{all},\mathrm{RF}}}(n)={{\left[ \mathbf{u}_{\mathrm{RF},1}^{\mathrm{T}}(n),\cdots ,\mathbf{u}_{\mathrm{RF},\frac{M}{{{M}_{\mathrm{RF}}}}}^{\mathrm{T}}(n) \right]}^{\mathrm{T}}}\in {{\mathbb{C}}^{ M\times 1}}$.
Then \eqref{yrfd} can be compactly written as
\begin{equation}
{{{\mathbf{y}}_{\mathrm{all},\mathrm{RF}}}(n)=\sqrt{{{P}_{d}}}\mathcal{V}^{\mathrm{H}}\mathbf{h}s\left( n \right)+{{\mathbf{u}}_{\mathrm{all},\mathrm{RF}}}(n)}.
\end{equation}
By applying the inverse matrix of $\mathcal{V}$ at the BS, we have
\begin{equation}
{{{\mathbf{y}}_{\mathrm{all}}}(n)={{\left( {{\mathcal{V}}^{\mathrm{H}}} \right)}^{-1}}{{\mathbf{y}}_{\mathrm{all},\mathrm{RF}}}(n)=\sqrt{{{P}_{d}}}\mathbf{h}s\left( n \right)+{{\mathbf{u}}_{\mathrm{all}}}(n)},\label{inverse}
\end{equation}
where ${{\mathbf{u}}_{\mathrm{all}}}(n)={{\left( {{\mathcal{V}}^{\mathrm{H}}} \right)}^{-1}}{{\mathbf{u}}_{\mathrm{all},\mathrm{RF}}}(n)$,
and the second last equality holds when $\mathcal{V}$ is full rank.
In order to meet the full rank requirements, the DFT codebook $\mathcal{V}$ can be selected as
\begin{equation}
{\mathcal{V}=\frac{1}{\sqrt{M}}\left[ \begin{matrix}
   1 & 1 & \cdots  & 1  \\
   1 & {{e}^{j2\pi \frac{1}{M}}} & \cdots  & {{e}^{j2\pi \frac{M-1}{M}}}  \\
   \vdots  & \vdots  & \ddots  & \vdots   \\
   1 & {{e}^{j2\pi \frac{M-1}{M}}} & \cdots  & {{e}^{j2\pi \frac{{{\left( M-1 \right)}^{2}}}{M}}}  \\
\end{matrix} \right]}\label{DFTcodebook}.
\end{equation}
Denote by $\mathbf{R}_{y}$ the covariance of the signals of $\mathbf{y}_{\mathrm{all}}(n)$ in \eqref{inverse}.
It can be approximated by taking the sample average of the received signal multiplied by its conjugate transpose
\begin{equation}
{{{\mathbf{R}}_{y}}\approx\frac{1}{\kappa }\sum\limits_{n=1}^{\kappa }{\mathbf{y}_{\mathrm{all}}(n)\mathbf{y}_{\mathrm{all}}^{\mathrm{H}}(n)}}.
\end{equation}
Then the angle can be estimated as $\hat{\theta }$ via conducting the one-dimensional spectral search of the Bartlett pseudospectrum to maximize ${{{P}_{\mathrm{Bartlett}}}\left( \theta  \right)={{\mathbf{a}}^{\mathrm{H}}}\left( \theta  \right){{\mathbf{R}}_{y}}\mathbf{a}\left( \theta  \right)}$.

On the other hand, if the data symbols $s\left(n \right)$ vary with different analog applications,
we use the digital covariance matrix reconstruction method in \cite{BSA}.
To achieve beam sweeping, the DFT codebook $\mathcal{V}$ in \eqref{DFTcodebook} can also be used here.

Then the resulting signal at the BS of the $i$-th search after analog beamforming during data symbol $s\left(n+i \right)$ is written as
\begin{equation}
\begin{split}
{{\mathbf{y}}_{\mathrm{RF}}}(n+i)& =\mathbf{V}_{\mathrm{RF},i}^{\mathrm{H}}{{\mathbf{y}}_{d}}(n+i)\\
& =\sqrt{{{P}_{d}}}\mathbf{V}_{\mathrm{RF},i}^{\mathrm{H}}\mathbf{h}s\left( n+i \right)+{{\mathbf{z}}_{\mathrm{RF},d}}\left( n+i \right).\label{ydifferents}
\end{split}
\end{equation}
In order to employ MUSIC algorithm and spatial smoothing for coherent signals, we reconstruct
the covariance matrix by BSA in \cite{BSA}.
Denote the covariance of ${{\mathbf{y}}_{\mathrm{RF}}}(n+i)$ in \eqref{ydifferents} as
${\mathbf{R}}_{\mathrm{RF},i}$.
When the sample size is sufficiently large, ${\mathbf{R}}_{\mathrm{RF},i}$ is represented as
\begin{equation}
\begin{split}
  {{\mathbf{R}}_{\mathrm{RF},i}}& \approx \frac{1}{\kappa }\sum\limits_{k=1}^{\kappa }{{{\mathbf{y}}_{\mathrm{RF}}}\left( k+i \right)\mathbf{y}_{\mathrm{RF}}^{\mathrm{H}}\left( k+i \right)} \\
 & =\mathbf{V}_{\mathrm{RF},i}^{\mathrm{H}}\frac{1}{\kappa }\sum\limits_{k=1}^{\kappa }{{{\mathbf{y}}_{d}}\left( k+i \right)\mathbf{y}_{d}^{\mathrm{H}}\left( k+i \right)}{{\mathbf{V}}_{\mathrm{RF},i}} \\
 & \approx \mathbf{V}_{\mathrm{RF},i}^{\mathrm{H}}{{\mathbf{R}}_{d}}{{\mathbf{V}}_{\mathrm{RF},i}},\label{Rrfi}
\end{split}
\end{equation}
where ${{\mathbf{R}}_{d}}$ is the covariance matrix of the received signals before analog beamforming ${\mathbf{y}}_{d}(n)$.
In order to reconstruct ${{\mathbf{R}}_{d}}$ from ${\mathbf{R}}_{\mathrm{RF},i}$, we first vectorize \eqref{Rrfi} as
\begin{equation}
\begin{split}
  {{\mathbf{c}}_{i}}& =\mathrm{vec}\left( \mathbf{V}_{\mathrm{RF},i}^{\mathrm{H}}{{\mathbf{R}}_{d}}{{\mathbf{V}}_{\mathrm{RF},i}} \right) \\
 & =\left( \mathbf{V}_{\mathrm{RF},i}^{\mathrm{T}}\otimes \mathbf{V}_{\mathrm{RF},i}^{\mathrm{H}} \right)\mathrm{vec}\left( {{\mathbf{R}}_{d}} \right) \\
 & ={{\left( {{\mathbf{V}}_{\mathrm{RF},i}}\otimes \mathbf{V}_{\mathrm{RF},i}^{*} \right)}^{\mathrm{T}}}{{\mathbf{r}}_{d}}, \label{ci}
\end{split}
\end{equation}
where ${{\mathbf{r}}_{d}}=\mathrm{vec}\left( {{\mathbf{R}}_{d}} \right)$,
and the second last equality holds for the identity that $\mathrm{vec}\left( \mathbf{ABC} \right)=\left( {{\mathbf{C}}^{\mathrm{T}}}\otimes \mathbf{A} \right)\mathrm{vec}\left( \mathbf{B} \right)$.
Concatenating ${{\mathbf{c}}_{i}}$ in \eqref{ci} for all $i = 1,\cdots, \frac{M}{M_{\mathrm{RF}}}$ as
$\mathbf{c}={{\big[ \mathbf{c}_{1}^{\mathrm{T}},\cdots ,\mathbf{c}_{\frac{M}{{{M}_{\mathrm{RF}}}}}^{\mathrm{T}} \big]}^{\mathrm{T}}}$,
equation \eqref{ci} can be further extended as
\begin{equation}
{\mathbf{c}=\mathbf{V}{{\mathbf{r}}_{d}}},
\end{equation}
where
\begin{equation}
{\mathbf{V}=\left[ \begin{matrix}
   {{\left( {{\mathbf{V}}_{\mathrm{RF},1}}\otimes \mathbf{V}_{\mathrm{RF},1}^{*} \right)}^{\mathrm{T}}}  \\
   \vdots   \\
   {{\left( {{\mathbf{V}}_{\mathrm{RF},\frac{M}{{{M}_{\mathrm{RF}}}}}}\otimes \mathbf{V}_{\mathrm{RF},\frac{M}{{{M}_{\mathrm{RF}}}}}^{*} \right)}^{\mathrm{T}}}  \\
\end{matrix} \right]\in {\mathbb{C}}^{{{M}_{\mathrm{RF}}}M\times {{M}^{2}}}}.\label{Vequation}
\end{equation}
Since $\mathbf{V}$ may not be a full rank matrix,
direct calculation of \eqref{Vequation} may be hampered by the ill-conditioned solution.
Therefore, diagonal loading can be utilized here to improve the distribution of the eigenvalues and handle the ill-conditioned issue \cite{illsolution}.
Denote the diagonal loading coefficient as $\delta$.
Then the vector ${\mathbf{r}_{d}}$ can be solved as
\begin{equation}
{{{\hat{\mathbf{r}}}_{d}}={{\left( {{\mathbf{V}}^{\mathrm{H}}}\mathbf{V}+\delta {{\mathbf{I}}_{{{M}^{2}}}} \right)}^{-1}}{{\mathbf{V}}^{\mathrm{H}}}\mathbf{c}}.\label{rdcal}
\end{equation}
Finally, the desired spatial covariance matrix can be reconstructed through
\begin{equation}
{{{\mathbf{\hat{\mathbf{R}}}}_{d}}=\mathrm{unvec}\left( {{{\hat{\mathbf{r}}}}_{d}} \right)},
\end{equation}
where unvec is the reverse operation of vec.
To greatly reduce the substantial computational burden that matrix inversion triggers,
the operator ${{\left( {{\mathbf{V}}^{\mathrm{H}}}\mathbf{V}+\delta {{\mathbf{I}}_{{{M}^{2}}}} \right)}^{-1}}{{\mathbf{V}}^{\mathrm{H}}}$ in \eqref{rdcal} can be pre-calculated off-line \cite{BSA}.
With the obtained spatial covariance matrix, the angle can be estimated as $\hat{\theta }$ via conducting the one-dimensional spectral search of the Bartlett pseudospectrum to maximize ${{{P}_{\mathrm{Bartlett}}}\left( \theta  \right)={{\mathbf{a}}^{\mathrm{H}}}\left( \theta  \right){{\mathbf{\hat{\mathbf{R}}}}_{d}}\mathbf{a}\left( \theta  \right)}$.

After the angle has been estimated, the hybrid receive beamforming vector ${{\mathbf{v}}_{\mathrm{eff}}}={{\mathbf{v}}_{\mathrm{RF}}}{{v}_{\mathrm{BB}}}=\frac{\mathbf{a}\left( {\hat{\theta }} \right)}{\left\| \mathbf{a}\left( {\hat{\theta }} \right) \right\|}$ is performed at the BS to detect communication symbols, where ${{\mathbf{v}}_{\mathrm{RF}}}=\frac{\mathbf{a}\left( {\hat{\theta }} \right)}{\left\| \mathbf{a}\left( {\hat{\theta }} \right) \right\|}\in {{\mathbb{C}}^{M\times 1}}$ and ${v_{\mathrm{BB}}}=1$.
That is to say, the number of RF chains ${M_{\mathrm{RF}}}=1$.
Then the resulting signal after beamforming is written as
\begin{equation}
\begin{split}
{y_{d,\mathrm{eff}}(n)}& =\mathbf{v}_{\mathrm{eff}}^{\mathrm{H}}{{\mathbf{y}}_{d}(n)}\\
& =\sqrt{{{P}_{d}}}\mathbf{v}_{\mathrm{eff}}^{\mathrm{H}}\mathbf{h}{s}\left( n \right)+{u}_{d,\mathrm{eff}}\left( n \right),\label{afterbeamangleL1}
\end{split}
\end{equation}
in which ${u}_{d,\mathrm{eff}}\left( n \right)=\mathbf{v}_{\mathrm{eff}}^{\mathrm{H}}{{\mathbf{u}}_{d}\left( n \right)}$ is the CSCG noise with zero-mean and variance ${\sigma }^{2}$.

\emph{Theorem 2:}
The expected SNR of the signal in \eqref{afterbeamangleL1} is written as
\begin{equation}
	{{{\gamma }_{\mathrm{eff}}}=\mathbb{E}\Big[ \frac{{{P}_{d}}{{\left| \mathbf{v}_{\mathrm{eff}}^{\mathrm{H}}{\bf h} \right|}^{2}}}{{{\sigma }^{2}}} \Big]=\frac{{{P}_{d}}{{\left\| {\bf h} \right\|}^{2}}}{{{\sigma }^{2}}}\frac{{{\left| {{\mathbf{a}}^{\mathrm{H}}}\big( {\hat{\theta }} \big)\mathbf{a}\left( \theta \right) \right|}^{2}}}{M^{2}}}.\label{gamma2initial}
\end{equation}
\begin{IEEEproof}
By substituting ${{\mathbf{v}}_{\mathrm{eff}}}=\frac{\mathbf{a}\left( {\hat{\theta }} \right)}{\sqrt{M}}$ into (\ref{gamma2initial}), we have
\begin{equation}
\begin{split}
  {{\gamma }_{\mathrm{eff}}}& =\mathbb{E}\Big[ \frac{{{P}_{d}}{{\left| \mathbf{v}_{\mathrm{eff}}^{\mathrm{H}}\mathbf{h} \right|}^{2}}}{{{\sigma }^{2}}} \Big]=\frac{{{P}_{d}}}{M{{\sigma }^{2}}}{{\left| {{\mathbf{a}}^{\mathrm{H}}}\left( {\hat{\theta }} \right)\mathbf{h} \right|}^{2}} \\
 & =\frac{{{P}_{d}}{{\left| \alpha  \right|}^{2}}{{\left\| \mathbf{a}\left( \theta  \right) \right\|}^{2}}}{{{\sigma }^{2}}}\frac{{{\left| {{\mathbf{a}}^{\mathrm{H}}}\left( {\hat{\theta }} \right)\mathbf{a}\left( \theta  \right) \right|}^{2}}}{M{{\left\| \mathbf{a}\left( \theta  \right) \right\|}^{2}}} \\
 & =\frac{{{P}_{d}}{{\left\| \mathbf{h} \right\|}^{2}}}{{{\sigma }^{2}}}\frac{{{\left| {{\mathbf{a}}^{\mathrm{H}}}\left( {\hat{\theta }} \right)\mathbf{a}\left( \theta  \right) \right|}^{2}}}{{{M}^{2}}},
\end{split}
\end{equation}
then Theorem 2 is proved.
\end{IEEEproof}
Theorem 2 indicates that the expected receive SNR is related to the performance of the angle estimation.
If the angle is estimated without error, i.e., ${\hat{\theta }} = \theta$,
there is ${{\gamma }_{\mathrm{eff}}}=\frac{{{P}_{d}}{{\left\| \mathbf{h} \right\|}^{2}}}{{{\sigma }^{2}}}.$
That is to say, with the accurately estimated angle, the expected SNR is able to achieve the upper bound ${{\gamma }_{\mathrm{upper}}}$ in \eqref{upper}.

Next, to detect the communication symbol, we need to further estimate the channel path gain.
Little pilots need to be utilized with the implement of the hybrid receive beamforming ${{\mathbf{v}}_{\mathrm{eff}}}$ towards $\hat{\theta }$ at the BS as
\begin{equation}
\begin{split}
y_{t,\mathrm{eff}}(n)& =\frac{{{\mathbf{a}}^{\mathrm{H}}}\big( {\hat{\theta }} \big)}{\big\| \mathbf{a}\big( {\hat{\theta }} \big) \big\|}{{\mathbf{y}_{t}(n)}}\\
& =\sqrt{{{P}_{t}}}\alpha \frac{{{\mathbf{a}}^{\mathrm{H}}}\big( {\hat{\theta }} \big)}{\big\| \mathbf{a}\big( {\hat{\theta }} \big) \big\|}\mathbf{a}\left( \theta  \right)\phi(n)+u_{t,\mathrm{eff}}(n),
\end{split}
\end{equation}
where $u_{t,\mathrm{eff}}(n)=\frac{{{\mathbf{a}}^{\mathrm{H}}}\left( {\hat{\theta }} \right)}{\left\| \mathbf{a}\left( {\hat{\theta }} \right) \right\|}\mathbf{u}_{t}(n).$
Projecting $y_{t,\mathrm{eff}}(n)$ to the known pilot symbol $\phi(n)$, we have
\begin{equation}
\begin{split}
{{y}_{t,\mathrm{eff}}^{\prime }}& =\frac{1}{\sqrt{{{P}_{t}}{{\rho }^{2}}}}\sum\limits_{n=1}^{\rho }{{{y}_{t,\mathrm{eff}}}\left( n \right)\phi^{*} \left( n \right)}\\
& =\alpha \frac{{{\mathbf{a}}^{\mathrm{H}}}\big( {\hat{\theta }} \big)}{\big\| \mathbf{a}\big( {\hat{\theta }} \big) \big\|}\mathbf{a}\left( \theta  \right)+\frac{1}{\sqrt{{{P}_{t}}{{\rho }^{2}}}}{{u}_{t,\mathrm{eff}}^{\prime }},
\end{split}
\end{equation}
where $u_{t,\mathrm{eff}}^{\prime }=\sum\limits_{n=1}^{\rho }{{{u}_{t,\mathrm{eff}}}\left( n \right)}{{\phi }^{*}}\left( n \right)$ denotes the noise vector, satisfying
$u_{t,\mathrm{eff}}^{\prime }\sim\mathcal{C}\mathcal{N}\left( 0, {{\rho }}{{\sigma }^{2}}\right)$.
Then $\alpha$ can be estimated by LS estimation scheme as
\begin{equation}
{\hat \alpha = \frac{{\left\| {{\bf{a}}(\hat \theta )} \right\|{y_{t,\mathrm{eff}}^{\prime }} }}{{{{\bf{a}}^{\rm{H}}}(\hat \theta ){\bf{a}}\left( \theta  \right)}}
\mathop  \approx \limits^{(a)} \frac{{{y_{t,\mathrm{eff}}^{\prime }} }}{{\sqrt M }}},\label{alphaes}
\end{equation}
in which $(a)$ holds under the assumption of ${\hat{\theta }} \approx \theta$.
Further, the estimation error of ${\mathbf{h}}$ is written as
\begin{equation}
\begin{split}
  {{{\mathbf{\tilde{h}}}}_{\mathrm{eff}}}&=\hat{\alpha }\mathbf{a}\big( {\hat{\theta }} \big)-\alpha \mathbf{a}\left( \theta  \right)\\
  & =\frac{{{y}_{t,\mathrm{eff}}^{\prime }}}{\sqrt{M}}\mathbf{a}\big( {\hat{\theta }} \big)-\alpha \mathbf{a}\left( \theta  \right) \\
 & =\frac{\alpha }{M}{{\mathbf{a}}^{\mathrm{H}}}\big( {\hat{\theta }} \big)\mathbf{a}\left( \theta  \right)\mathbf{a}\big( {\hat{\theta }} \big)-\alpha \mathbf{a}\left( \theta  \right)+\frac{1}{\sqrt{{{P}_{t}}{{\rho }^{2}}M}}{{u}_{t,\mathrm{eff}}^{\prime }}\mathbf{a}\big( {\hat{\theta }} \big)\\
 & \mathop  \approx \limits^{(a)}  \frac{1}{\sqrt{{{P}_{t}}{{\rho }^{2}}M}}{{u}_{t,\mathrm{eff}}^{\prime }}\mathbf{a}\big( {\hat{\theta }} \big).\label{tildeh2}
\end{split}
\end{equation}
Then the MMSE of channel ${\mathbf{h}}$ is written as
\begin{equation}
\begin{split}
e_\mathrm{eff}& =\mathbb{E}\big[ {{\big\| {{{\mathbf{\tilde{h}}}}_{\mathrm{eff}}} \big\|}^{2}} \big]\\
& ={{\alpha }^{2}}M-\frac{{{\alpha }^{2}}}{M}{{\left| {{\mathbf{a}}^{\mathrm{H}}}\left( \theta  \right)\mathbf{a}\big( {\hat{\theta }} \big) \right|}^{2}}+\frac{{{\sigma }^{2}}}{{{P}_{t}}\rho }
\mathop \approx \limits^{(a)}\frac{{{\sigma }^{2}}}{{{P}_{t}}\rho }.\label{RMSEe2}
\end{split}
\end{equation}
Compared with $e_\mathrm{con}$ in \eqref{RMSEe1}, it is observed that our proposed method is about to achieve $M$ times more accurate channel estimation, compared with the conventional scheme.

\subsection{Multipath Channel}
Next, the more general case with multipath channel is studied, i.e., $L>1$.
We first consider the fully digital structure at the BS to obtain some insights about
the performance of estimation and signal detection.
\subsubsection{Fully Digital Structure}
Based on the fact that subspace-based algorithms like MUSIC only works effectively when multiple propagation paths are independent, the spatial smoothing technique is used here to tackle the coherence sources of different directions \cite{FBSS}.
Divide the antenna array of the BS into $G$ overlapping subarrays, then the number of elements for every subarray is $M_{\mathrm{sub}}=M-G+1$.
The output of the $g$-th forward subarray in pilot transmission phase in \eqref{originalpilotphase} is formulated as
\begin{equation}
\begin{split}
  {{\mathbf{y}}_{t,g}(n)}& =\left[ {y_{t,g}(n)},\cdots ,{y_{t,\left( g+{{M}_{\mathrm{sub}}}-1 \right)}}(n) \right]^{\mathrm{T}}\\
  & ={{\mathbf{A}}_{g}\left(\Theta  \right)}\bm{\alpha}\sqrt{{{P}_{t}}}{{\phi}(n)}+\mathbf{u}_{t,g}(n),\label{subequation}
\end{split}
\end{equation}
where ${y_{t,g}(n)}$ is the $g$-th element of ${\mathbf{y}_{t}(n)}$.
${{\mathbf{u}}_{t,g}(n)}=\left[ {u}_{t,g}(n),\cdots ,{u_{t,\left( g+{{M}_{\mathrm{sub}}}-1 \right)}(n)} \right]^{\mathrm{T}}\in {{\mathbb{C}}^{{{M}_{\mathrm{sub}}}\times 1}}$, $g=1,...,G$.
${{y}_{t,m}(n)}$ and ${{u}_{t,m}(n)}, m=1,...,M$ is the $m$ element of ${\mathbf{y}}_{t}(n)$ and ${\mathbf{u}}_{t}(n)$, respectively.
In addition, we have ${{\mathbf{A}}_{g}\left(\bm{\Theta}  \right)}=\left[ {{\mathbf{a}}_{g}}\left( {{\theta }_{1}} \right),\cdots ,{{\mathbf{a}}_{g}}\left( {{\theta }_{L}} \right) \right]\in {{\mathbb{C}}^{{{M}_{\mathrm{sub}}}\times L}}$,
where ${{\mathbf{a}}_{g}}\left( \theta_{l}  \right)={{\left[ {{e}^{j\pi g\sin \left( \theta_{l}  \right)}},\cdots ,{{e}^{j\pi \left( g+{{M}_{\mathrm{sub}}}-1 \right)\sin \left( \theta_{l}  \right)}} \right]}^{\mathrm{T}}}\in {{\mathbb{C}}^{{{M}_{\mathrm{sub}}}\times 1}}$.
Let ${\mathbf{B}=\mathrm{diag}\left[ {{e}^{-j\pi \sin {{\theta }_{1}}}},\cdots ,{{e}^{-j\pi \sin {{\theta }_{L}}}} \right]},$ then \eqref{subequation} can also be expressed as
\begin{equation}
{{{\mathbf{y}}_{t,g}(n)}={{\mathbf{A}}_{1}\left(\bm{\Theta}  \right)}{{\mathbf{B}}^{g-1}}\bm{\alpha}\sqrt{{{P}_{t}}}{{\phi}(n)}+\mathbf{u}_{t,g}(n).}\label{y1p}
\end{equation}
Similarly, the output of the $g$-th forward subarray during data transmission in \eqref{originaldataphase} can be expressed as
\begin{equation}
{{{\mathbf{y}}_{d,g}(n)}={{\mathbf{A}}_{1}\left(\bm{\Theta}  \right)}{{\mathbf{B}}^{g-1}}\bm{\alpha}\sqrt{{{P}_{d}}}{s(n)}+\mathbf{u}_{d,g}(n).}\label{y2p}
\end{equation}
Denote by $\mathbf{R}_{g}$ the covariance of $g$-th subarray in \eqref{y1p} and \eqref{y2p},
which can be approximated by taking the sample average of the received signal multiplied by its conjugate transpose
\begin{equation}
{{{\mathbf{R}}_{g}}\approx\frac{1}{\rho +\kappa }(\sum\limits_{n=1}^{\rho }{{{\mathbf{y}}_{t,g}}\left( n \right)\mathbf{y}_{t,g}^{\mathrm{H}}\left( n \right)}+\sum\limits_{n=\rho+1}^{\rho+\kappa }{{{\mathbf{y}}_{d,g}}\left( n \right)\mathbf{y}_{d,g}^{\mathrm{H}}\left( n \right)})}.
\end{equation}
Based on the average of the covariance matrices of all $G$ subarrays, the forward and backward spatial smoothing covariance matrix can be respectively written as
\begin{equation}
{\mathbf{R}^{\mathrm{forward}}=\frac{1}{G}\sum\limits_{g=1}^{G}{{{\mathbf{R}}_{g}}}}\label{forcov},
\end{equation}
and
\begin{equation}
{{{\mathbf{R}}^{\mathrm{backward}}}=\frac{1}{G}\sum\limits_{g=1}^{G}{{{{\mathbf{\tilde{R}}}}_{g}}}},
\end{equation}
where ${{\mathbf{\tilde{R}}}_{g}}={{\mathbf{Q}}_{{{M}_{\mathrm{sub}}}}}\mathbf{R}_{g}^{*}{{\mathbf{Q}}_{{{M}_{\mathrm{sub}}}}}$,
and ${{\mathbf{Q}}_{{{M}_{\mathrm{sub}}}}}$ is a ${{M}_{\mathrm{sub}}}$-order exchange matrix with zero except for the elements on the subdiagonal, which are 1.
Then the bidirectional spatial smoothing covariance matrix is obtained \cite{FBSS}
\begin{equation}
{\mathbf{R}_{\mathrm{bi}}=\frac{1}{2}\left( {{\mathbf{R}}^{\mathrm{forward}}}+{{\mathbf{R}}^{\mathrm{backward}}} \right)}\label{allcov}.
\end{equation}
If the number of subarrays $G$ and elements of each subarray ${{M}_{\mathrm{sub}}}$ satisfy ${{M}_{\mathrm{sub}}} \ge L+1$ and $2G\ge L$, respectively,
the bidirectional spatial smoothing covariance matrix $\mathbf{R}_{\mathrm{bi}}$ is full rank.
In this case, the angles of multipath channel components can be estimated
by standard MUSIC algorithm.

In the following, the complex-valued path coefficients $\bm{\alpha}$ are further estimated to detect the communication symbol, where the direction of $L$ multipaths should be considered when constructing the receive beamforming matrix.
Specifically, in order to match the $l$-th path,
the receive beamforming vector $\frac{\mathbf{a}\left( {{\hat{\theta} }_{l}} \right)}{\left\| \mathbf{a}\left( {{\hat{\theta} }_{l}} \right) \right\|}$ can be employed with the pilot sequence $\bm{\phi}_{l}$, which results in
\begin{equation}
  {\mathbf{y}_{t,\left( l \right)}^{\mathrm{H}}=\frac{\sqrt{{{P}_{t}}}}{\sqrt{M}}{{\mathbf{a}}^{\mathrm{H}}}\left( {{{\hat{\theta }}}_{l}} \right)\mathbf{h}\bm{\phi }_{l}^{\mathrm{H}}+\frac{1}{\sqrt{M}}{{\mathbf{a}}^{\mathrm{H}}}\left( {{{\hat{\theta }}}_{l}} \right){{\mathbf{U}}_{t}}}.
\end{equation}
Similar to the analysis in Section \ref{los}, project $\mathbf{y}_{t,\left( l \right)}^{\mathrm{H}}$ to the known pilot sequence $\bm{\phi}_{l}$, we have
\begin{equation}
{y_{t,\left( l \right)}=\frac{1}{\rho }\mathbf{y}_{t,\left( l \right)}^{\mathrm{H}}{{\bm{\phi }}_{l}}=\frac{\sqrt{{{P}_{t}}}}{\sqrt{M}}{{\mathbf{a}}^{\mathrm{H}}}\left( {{{\hat{\theta }}}_{l}} \right)\mathbf{h}+\frac{1}{\sqrt{M{{\rho }^{2}}}}u_{t,\left( l \right)},}\label{ylesalpha}
\end{equation}
where $u_{t,\left( l \right)}={{\mathbf{a}}^{\mathrm{H}}}\left( {{{\hat{\theta }}}_{l}} \right){{\mathbf{U}}_{t}}{{\bm{\phi }}_{l}}$ is i.i.d. CSCG noise with power $M\rho{\sigma }^{2}$.
By concatenating $y_{t,\left( l \right)}$ in \eqref{ylesalpha} for all $l = 1, ..., L$, we have
\begin{equation}
\begin{split}
  {{\mathbf{y}}_{t,\mathrm{mp}}}& ={{\left[ y_{t,\left( l \right)},\cdots ,y_{t,\left( l \right)} \right]}^{\mathrm{T}}} \\
 & =\frac{\sqrt{{{P}_{t}}}}{\sqrt{M}}{{\mathbf{A}}^{\mathrm{H}}}(\mathbf{\hat{\Theta }})\mathbf{A}(\mathbf{\Theta})\bm{\alpha}+\frac{1}{\sqrt{M{{\rho }^{2}}}}{{\mathbf{u}}_{t,\mathrm{mp}}},\label{ytMP}
\end{split}
\end{equation}
where ${{\mathbf{u}}_{t,\mathrm{mp}}}={{\left[ u_{t,\left( l \right)},\cdots ,u_{t,\left( l \right)} \right]}^{\mathrm{T}}}$ denotes i.i.d. CSCG noise with power $M\rho{\sigma }^{2}$.
$\bm{\alpha}$ can be estimated by LS method as
\begin{equation}
\begin{split}
\bm{\hat{\alpha }}& = \frac{{\sqrt M }}{{\sqrt {{P_t}} }}{({{\bf{A}}^{\rm{H}}}({\bf{\hat \Theta }}){\bf{A}}({\bf{\Theta }}))^{ - 1}}{{\bf{y}}_{t,{\rm{mp}}}}\\
& \mathop\approx \limits^{(b)} \frac{{\sqrt M }}{{\sqrt {{P_t}} }}{({{\bf{A}}^{\rm{H}}}({\bf{\hat \Theta }}){\bf{A}}({\bf{\hat \Theta }}))^{ - 1}}{{\bf{y}}_{t,{\rm{mp}}}}
,\label{bmalpha}
\end{split}
\end{equation}
where $(b)$ holds due to ${\mathbf{\Theta }}\approx{\mathbf{\hat{\Theta }}}$.
Then the estimation of $\mathbf{h}_{\mathrm{eff}}$ is ${{\mathbf{\hat{h}}}_{\mathrm{eff}}}=\mathbf{A}\left( {\mathbf{\hat{\Theta }}} \right)\bm{\hat{\alpha }}$,
and the estimation error of ${{\mathbf{{h}}}_{\mathrm{eff}}}$ is formulated as
\begin{equation}
\begin{split}
  & {{{\mathbf{\tilde{h}}}}_{\mathrm{eff}}}={{{\mathbf{\hat{h}}}}_{\mathrm{eff}}}-\mathbf{h}_{\mathrm{eff}}=\mathbf{A}\big( {\mathbf{\hat{\Theta }}} \big)\bm{\hat{\alpha} }-\mathbf{A}\left( \mathbf{\Theta } \right)\bm{\alpha }  \\
 & =\frac{\sqrt{M}}{\sqrt{{{P}_{t}}}}\mathbf{A}\big( {\mathbf{\hat{\Theta }}} \big){{\left( {{\mathbf{A}}^{\mathrm{H}}}\big( {\mathbf{\hat{\Theta }}} \big)\mathbf{A}\big( {\mathbf{\hat{\Theta }}} \big) \right)}^{-1}}{{\mathbf{y}}_{t,\mathrm{mp}}}-\mathbf{A}\big( \mathbf{\Theta } \big)\bm{\alpha }.\label{tildehulpmulti}
\end{split}
\end{equation}
When the angle is perfectly estimated, \eqref{tildehulpmulti} can be written as
\begin{equation}
{{{\mathbf{\tilde{h}}}_{\mathrm{eff}}}=\frac{1}{\sqrt{{{P}_{t}}{\rho}^{2}}}\mathbf{A}\big( {\mathbf{\hat{\Theta }}} \big){{\Big( {{\mathbf{A}}^{\mathrm{H}}}\big( {\mathbf{\hat{\Theta }}} \big)\mathbf{A}\big( {\mathbf{\hat{\Theta }}} \big) \Big)}^{-1}}{{\mathbf{u}}_{t,\mathrm{mp}}}}.
\end{equation}

\emph{Theorem 3:}
For multi-path communication with fully digital receive beamforming, the channel estimation accuracy of the proposed scheme can be evaluated by calculating the MMSE of ${\mathbf{h}}$ as
\begin{equation}
 {{{e}_{\mathrm{eff}}}=\mathbb{E}\big[ {{\big\| {{{\mathbf{\tilde{h}}}}_{\mathrm{eff}}} \big\|}^{2}} \big]=\frac{L{{\sigma }^{2}}}{{{P}_{t}\rho}}}.\label{elp}
\end{equation}
\begin{IEEEproof}
Please refer to Appendix B.
\end{IEEEproof}
Compared with $e_\mathrm{con}$ in \eqref{RMSEe1}, it is observed that when $L<M$,
the MMSE of the proposed efficient channel estimation method is smaller than that the conventional scheme.

Furthermore, to detect the communication symbol, the optimal receive beamforming vector
${{\mathbf{v}}_{\mathrm{eff}}}=\frac{{{{\mathbf{\hat{h}}}}_{\mathrm{eff}}}}{\left\| {{{\mathbf{\hat{h}}}}_{\mathrm{eff}}} \right\|}$ is considered, which results in
\begin{equation}
{{y_{d,\mathrm{eff}}(n)}=\mathbf{v}_{\mathrm{eff}}^{\mathrm{H}}{{\mathbf{y}}_{d}(n)}
=\sqrt{{{P}_{d}}}\mathbf{v}_{\mathrm{eff}}^{\mathrm{H}}\mathbf{h}{s}\left( n \right)+{u}_{d,\mathrm{eff}}\left( n \right)},
\end{equation}
where ${u}_{d,\mathrm{eff}}\left( n \right)=\mathbf{v}_{\mathrm{eff}}^{\mathrm{H}}{{\mathbf{u}}_{d}\left( n \right)}$.
The expected SNR of the signal in \eqref{afterbeamangleL1} is written as
\begin{equation}
  {{\gamma }_{\mathrm{eff}}}=\mathbb{E}\bigg[ \frac{{{P}_{d}}{{\big| \mathbf{v}_{\mathrm{eff}}^{\mathrm{H}}\mathbf{h} \big|}^{2}}}{{{\sigma }^{2}}} \bigg]=\frac{{{P}_{d}}}{{{\sigma }^{2}}}{{\mathbf{h}}^{\mathrm{H}}}\mathbb{E}\left[ {{\mathbf{v}}_{\mathrm{eff}}}\mathbf{v}_{\mathrm{eff}}^{\mathrm{H}} \right]\mathbf{h}.\label{gammaulpmulti}
\end{equation}
\emph{Theorem 4:}
The upper bound of ${{\gamma }_{\mathrm{eff}}}$ is
\begin{equation}
  {{{\gamma }_{\mathrm{eff}}} \le {{\gamma }_{\mathrm{upper}}}=\frac{{{P}_{d}}}{{{\sigma }^{2}}}{{\left\| \mathbf{h} \right\|}^{2}}}.\label{SNRmultipathUP}
\end{equation}
\begin{IEEEproof}
Please refer to Appendix C.
\end{IEEEproof}


\subsubsection{Hybrid Analog/digital Structure}

Next, we study the hybrid receive beamforming structure at the BS for more practical scenarios.
By passing through a hybrid beamforming preprocessor,
element space outputs at the BS is transformed into beamspace.
The digital signal reconstruction method and digital covariance matrix reconstruction method proposed in
Subsection III-A can be applied here to obtain the spatial covariance matrix.
Then the angles of multipath channel can be estimated accurately by MUSIC algorithm and forward-backward spatial smoothing technique in Section III-B 1).

To further estimate the complex-valued path coefficients $\bm{\alpha}$,
the receive beamforming vector $\frac{1}{\sqrt{M}}{\mathbf{A}}(\mathbf{\hat{\Theta }})\in {{\mathbb{C}}^{M\times L}}$ analyzed for the digital structure in \eqref{ytMP} can be directly used in the hybrid beamforming structure,
where $M_{\mathrm{RF}}=L$.

Furthermore, for the detection of the communication symbol, an analog beamforming scheme
that steers the beam towards the channel direction for $L$ paths is considered, i.e.,
${{\mathbf{V}}_{\mathrm{RF}}}=\frac{1}{\sqrt{M}}\mathbf{A}\left( {\mathbf{\hat{\Theta }}} \right)$.
The resulting signal after analog beamforming can be formulated as
\begin{equation}
{{\mathbf{y}_{d,\mathrm{RF}}}(n)=\mathbf{V}_{\mathrm{RF}}^{\mathrm{H}}{{\mathbf{y}}_{d}}(n)=\sqrt{{{P}_{d}}}\mathbf{V}_{\mathrm{RF}}^{\mathrm{H}}\mathbf{h}s\left( n \right)+{\mathbf{u}_{d,\mathrm{RF}}}\left( n \right)},\label{yRF}
\end{equation}
in which ${{\mathbf{u}}_{d,\mathrm{RF}}}\left( n \right)=\mathbf{V}_{\mathrm{RF}}^{\mathrm{H}}{{\mathbf{u}}_{d}}(n)$ is the resulting noise vector.
When the angles are precisely estimated, i.e., ${\mathbf{\Theta }}= {\mathbf{\hat{\Theta }}}$,
\eqref{yRF} can be expressed as
\begin{equation}
{{{\mathbf{y}}_{d,\mathrm{RF}}}(n)= \frac{\sqrt{{{P}_{d}}}}{\sqrt{M}}{{\mathbf{A}}^{\mathrm{H}}}\left( {\mathbf{\hat{\Theta }}} \right)\mathbf{A}\left( {\mathbf{\hat{\Theta }}} \right)\bm{\alpha }s\left( n \right)+{{\mathbf{u}}_{d,\mathrm{RF}}}\left( n \right)}.\label{yRFangle}
\end{equation}
In this case, the LS estimation of $\bm{\alpha }$ and the MMSE of $\mathbf{h}$
equal to $\hat{\bm{\alpha }}$ in \eqref{bmalpha} and $e_{\mathrm{eff}}$ in
\eqref{elp} in the fully digital structure.

According to \cite{zhengjiaoboshu}, if the azimuth AoAs in ULA system are independently generated from a continuous distribution,
as the number of antenna $M$ approaches infinity and that of the channel paths $L$ satisfies $L=o\left( M \right)$,
the receive array response vectors tends to orthogonal, i.e.,
$\mathbf{a}\left( {{\theta }_{l}} \right)\bot \mathrm{span}\left( \left\{ \mathbf{a}\left( {{\theta }_{k}} \right)|\forall k\ne l \right\} \right)$.
Therefore, ${{\mathbf{A}}^{\mathrm{H}}}\big( {\mathbf{\hat{\Theta }}} \big)\mathbf{A}\big( {\mathbf{\hat{\Theta }}} \big)\to M{{\mathbf{I}}_{L}}$
for the antennas of BS tends to infinity.
Then \eqref{yRFangle} is approximated as
\begin{equation}
{{{\mathbf{y}}_{d,\mathrm{RF}}}(n)\approx \sqrt{{{P}_{d}}M}\bm{\alpha }s\left( n \right)+{\mathbf{u}_{d,\mathrm{RF}}}\left( n \right)}.
\end{equation}
The receive digital beamforming matrix ${{\mathbf{v}}_{\mathrm{BB}}}=\frac{{{{\bm{\hat{\alpha }}}}}}{\left\| {\bm{\hat{\alpha }}} \right\|}$ can be used,
which results in
\begin{equation}
\begin{split}
  {{y}_{d,\mathrm{hybrid}}}(n)& =\mathbf{v}_{\mathrm{BB}}^{\mathrm{H}}{{\mathbf{y}}_{d,\mathrm{RF}}}(n) \\
 & =\frac{\sqrt{{{P}_{d}}M}}{\left\| {\bm{\hat{\alpha }}} \right\|}{{{\bm{\hat{\alpha }}}}^{\mathrm{H}}}\bm{\alpha }s\left( n \right)+{{u}_{d,\mathrm{hybrid}}}\left( n \right),\label{yhybrid}
\end{split}
\end{equation}
where ${{u}_{d,\mathrm{hybrid}}}\left( n \right)=\mathbf{v}_{\mathrm{BB}}^{\mathrm{H}}{{\mathbf{u}}_{d,\mathrm{RF}}}\left( n \right)$ is the CSCG noise with zero-mean and variance ${\sigma }^{2}$.

The expected SNR of the signal in \eqref{yhybrid} is written as
\begin{equation}
{{{\gamma }_{\mathrm{eff,hybrid}}}=\mathbb{E}\left[ \frac{{{P}_{d}}{{\left| \mathbf{v}_{\mathrm{BB}}^{\mathrm{H}}\mathbf{V}_{\mathrm{RF}}^{\mathrm{H}}\mathbf{h} \right|}^{2}}}{{{\sigma }^{2}}} \right]}.\label{gammahybridorigin}
\end{equation}
\emph{Theorem 5:}
\eqref{gammahybridorigin} can be approximated as
\begin{equation}
{{{\gamma }_{\mathrm{eff,hybrid}}}\approx \frac{{{P}_{d}}M{{\left\| \bm{\alpha } \right\|}^{2}}}{{{\sigma }^{2}}}\left( \frac{{{\left\| \bm{\alpha } \right\|}^{2}}+\frac{{{\sigma }^{2}}}{M{{P}_{t}}\rho }}{{{\left\| \bm{\alpha } \right\|}^{2}}+\frac{{{\sigma }^{2}}L}{M{{P}_{t}}\rho }} \right)}\label{gammalphybrid}.
\end{equation}
\begin{IEEEproof}
Please refer to Appendix D.
\end{IEEEproof}

Compared with the upper bound of the expected SNR in \eqref{SNRmultipathUP}, the SNR loss defined by ${{\gamma }_{\Delta }}$ is written as
\begin{equation}
\begin{split}
  {{\gamma }_{\Delta }}& ={{\gamma }_{\mathrm{upper}}}-{{\gamma }_{\mathrm{eff,hybrid}}} \\
 & =\frac{{{P}_{d}}{{\left\| \mathbf{h} \right\|}^{2}}}{{{\sigma }^{2}}}-\frac{{{P}_{d}}M{{\left\| \bm{\alpha } \right\|}^{2}}}{{{\sigma }^{2}}}\left( \frac{{{\left\| \bm{\alpha } \right\|}^{2}}+\frac{{{\sigma }^{2}}}{M{{P}_{t}}\rho }}{{{\left\| \bm{\alpha } \right\|}^{2}}+\frac{{{\sigma }^{2}}L}{M{{P}_{t}}\rho }} \right) \\
 & ={{P}_{d}}{{\left\| \bm{\alpha } \right\|}^{2}}\left( \frac{L-1}{\rho {{P}_{t}}{{\left\| \bm{\alpha } \right\|}^{2}}+\frac{{{\sigma }^{2}}L}{M}} \right),
\end{split}
\end{equation}
where the last equality holds for ${{\left\| \mathbf{h} \right\|}^{2}}\to M{{\left\| \bm{\alpha } \right\|}^{2}}$ under the assumption of ${{\mathbf{A}}^{\mathrm{H}}}\left( {\mathbf{\Theta }} \right)\mathbf{A}\left( {\mathbf{\Theta }} \right)\to M{{\mathbf{I}}_{L}}$.
It is shown that the loss of SNR can be reduced by increasing the length of pilots.
Furthermore, ${{\gamma }_{\Delta }}$ can also be written as
\begin{equation}
{{{\gamma }_{\Delta }}={{P}_{d}}{{\left\| \bm{\alpha } \right\|}^{2}}\left( \frac{M}{{{\sigma }^{2}}}-\frac{\frac{\rho M{{P}_{t}}{{\left\| \bm{\alpha } \right\|}^{2}}}{{{\sigma }^{2}}}+1}{\rho {{P}_{t}}{{\left\| \bm{\alpha } \right\|}^{2}}+\frac{{{\sigma }^{2}}L}{M}} \right)}.
\end{equation}
It is indicated that when the number of multipaths $L$ increases, the loss of SNR becomes larger.
When $L=M$, ${{\gamma }_{\Delta }}=\frac{{{P}_{d}}{{\left\| \bm{\alpha } \right\|}^{2}}M}{{{\sigma }^{2}}}\left( \frac{1-\frac{1}{M}}{\frac{\rho {{P}_{t}}{{\left\| \bm{\alpha } \right\|}^{2}}}{{{\sigma }^{2}}}+1} \right)$,
which equals to the penalty loss $\xi$ in \eqref{gamma1app}.
That is to say, when $L<M$, the receive SNR for our proposed scheme is
higher than that of the traditional scheme relying on pilot training.

\subsection{Pilot Overhead Comparison}
Finally, we compare the pilot overhead of our proposed method with the conventional method solely relying on channel training.
Let $\rho_{\mathrm{con}}$ and $\rho_{\mathrm{eff,hybrid}}$ denote the length of training sequence used by the conventional method in digital beamforming structure and the proposed method in hybrid beamforming structure, respectively.
To achieve the same performance for the expected SNR of the communication signal, i.e., $\gamma_{\mathrm{con}}$ in \eqref{gamma1app} is equal to $\gamma_{\mathrm{eff,hybrid}}$ in \eqref{gammalphybrid}, for $L>1$, we have
\begin{equation}
{{{\rho }_{\mathrm{con}}}=\frac{M-1}{L-1}{{\rho }_{\mathrm{eff,hybrid}}}+\frac{M-L}{M\mathrm{SNR}_{t}\left( L-1 \right)}}\label{pilotcompare}.
\end{equation}
That is to say, to achieve the same performance for the expected SNR, the conventional method needs to consume at least $\frac{M-1}{L-1}$ times pilots compared to the proposed method.

Furthermore, if the length of training sequence used by two methods are equal,
i.e., $\rho_{\mathrm{con}}=\rho_{\mathrm{eff,hybrid}}$,
the performance gap of the expected SNR between the proposed scheme and the traditional scheme can be written as
\begin{equation}
\begin{split}
{{\gamma }_{\mathrm{gap}}}& ={{\gamma }_{\mathrm{eff},\mathrm{hybrid}}}-{{\gamma }_{\mathrm{con}}}\\
& =M\mathrm{SNR}_{d}\frac{\left( 1-\frac{L}{M} \right)\left( \rho \mathrm{SNR}_{t}+\frac{1}{M} \right)}{\left( \rho \mathrm{SNR}_{t}+\frac{L}{M} \right)\left( \rho \mathrm{SNR}_{t}+1 \right)}.\label{twomethodsgap}
\end{split}
\end{equation}
With the assumption that the number of antenna $M$ approaches infinity and that of the channel paths satisfies $L=o\left( M \right)$, i.e., $M\to \infty $, $\frac{L}{M}\to 0$, ${{\gamma }_{\mathrm{gap}}}$ approaches to
\begin{equation}
{{{\gamma }_{\mathrm{gap}}}\to \frac{M\mathrm{SNR}_{d}}{\rho \mathrm{SNR}_{t}+1}}.\label{gapappro}
\end{equation}
It can be shown from \eqref{gapappro} that the performance gap increases with the number of antennas.
That is to say, compared with the conventional scheme, the proposed efficient channel estimation scheme enabled by ISSAC is expected to
obtain much better performance for large antenna systems operating at high frequency bands.
\section{Simulation Results}
In this section, simulation results are provided to compare
the performance of the proposed efficient channel estimation scheme enabled by ISSAC in the hybrid analog/digital beamforming structure
with the conventional scheme in the digital beamforming structure.
It is worth noting that the conventional scheme employing the digital beamforming structure can be seen as
the upper bound performance for that employing the hybrid analog/digital beamforming structure.
For mmWave and THz channels, we set the number of total multi-paths as $L=4$.
The path gains $\alpha_{l}$ are independently drawn from
$\alpha_{l}\sim\mathcal{C}\mathcal{N}\left( 0, 1\right), l=1,...,L$.
Define ${{\bar{P}}_{t}}=\frac{{{P}_{t}}}{{{\sigma }^{2}}}$ and
${{\bar{P}}_{d}}=\frac{{{P}_{d}}}{{{\sigma }^{2}}}$ as the transmit SNR by the UE during pilot and data symbol transmission phases, respectively.
Unless specified otherwise, we set ${\bar{P}}_{t}={\bar{P}}_{d}=-10$ dB, the length
of training sequence for two methods as $\rho=4$, the number of BS antennas as $M=64$,
and the number of RF chains as $M_{\mathrm{RF}}=L$.

\begin{figure}[!t]
  \centering
  \centerline{\includegraphics[width=3.1in,height=2.3in]{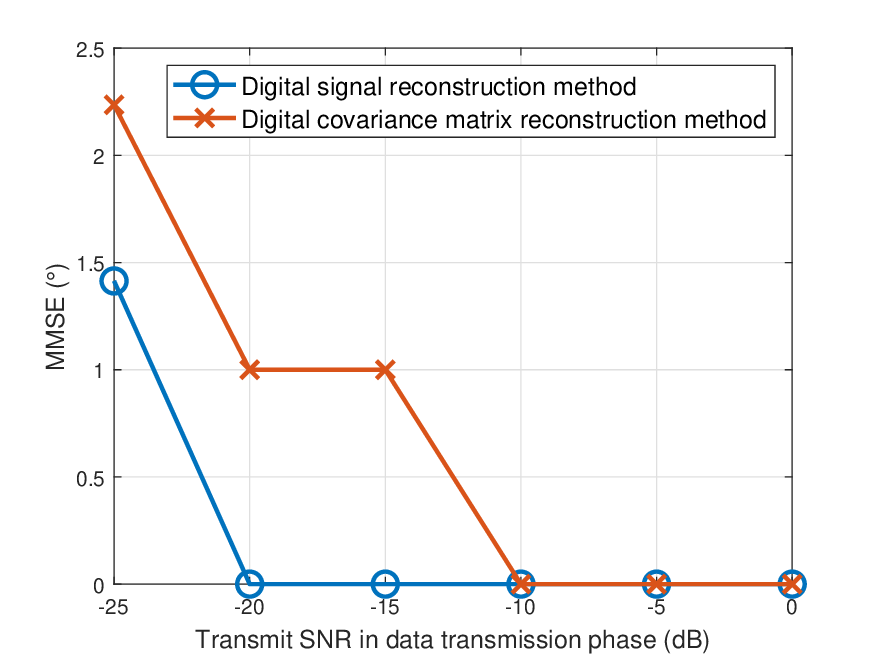}}
  \caption{MMSE of angles versus transmit SNR during data transmission phase ${\bar{P}}_{t}$.}
  \label{compareangleMMSE}
  \end{figure}
Fig. \ref{compareangleMMSE} plots the MMSE of the angles versus ${\bar{P}}_{d}$ for two methods introduced in Section III-B 2), i.e.,
the angle estimation method based on digital signal reconstruction and the method based on digital covariance matrix reconstruction.
It is observed from Fig. \ref{compareangleMMSE} that the estimation accuracy of both methods are consistent in most cases, such as when the SNR is more than -10 dB.
While the SNR is less than -10 dB, angle estimation scheme based on digital signal reconstruction achieves slightly better performance than the other one.
However, due to the involvement of high-dimensional matrix multiplication and inversion when solving linear equations,
the method based on digital covariance matrix reconstruction exhibits significantly high computational complexity.
Therefore, in the following, we use the estimation method based on digital signal reconstruction
to estimate AoAs.

\begin{figure}[htbp]
  \centering
    \begin{subfigure}{0.45\textwidth}
      \centering
      \includegraphics[width=1\linewidth]{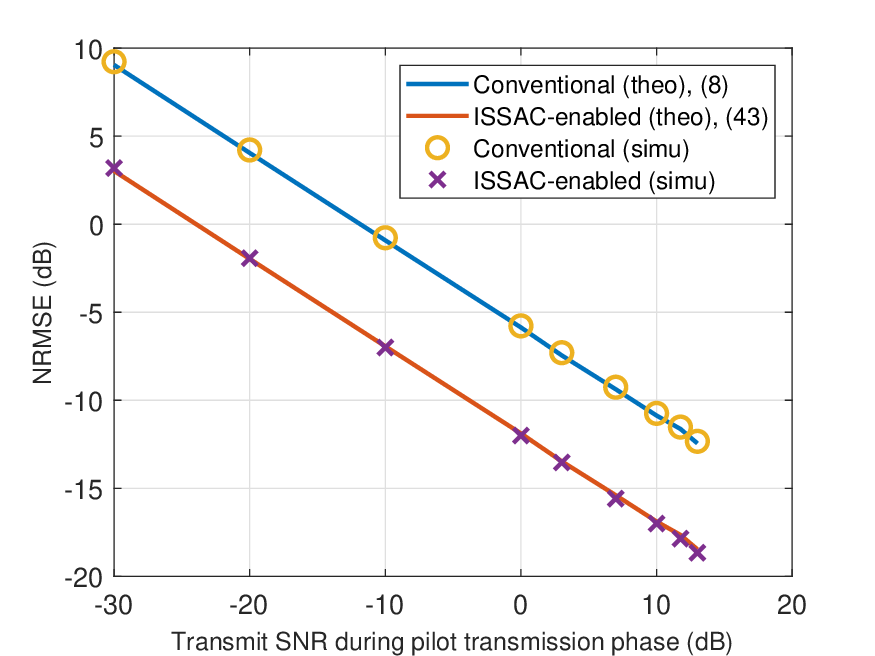}
        \caption{NRMSE versus transmit SNR during pilot training ${\bar{P}}_{t}$.}
        \label{NRMSEvsPt}
    \end{subfigure}   
     \begin{subfigure}{0.45\textwidth}
      \centering
      \includegraphics[width=1\linewidth]{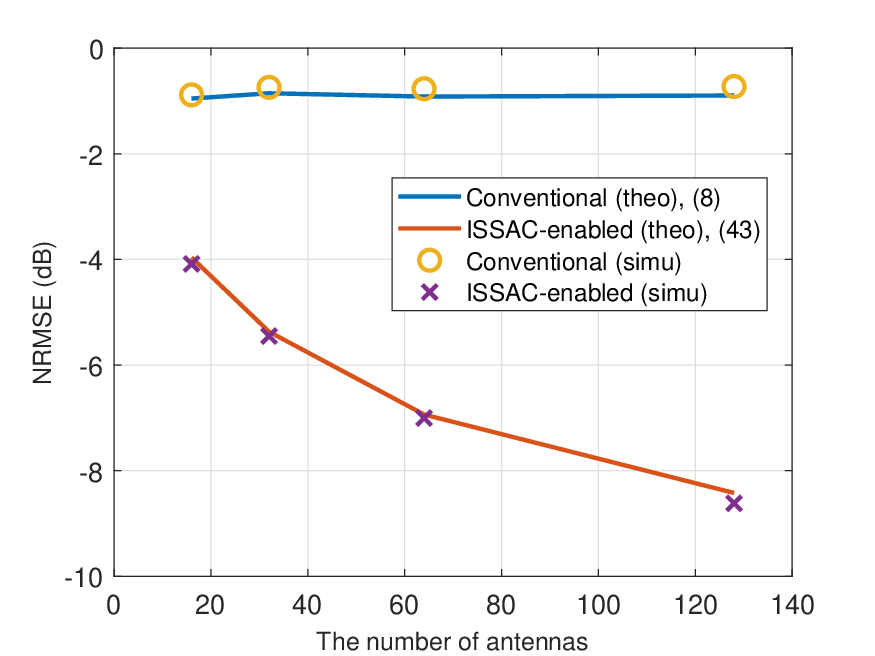}
        \caption{NRMSE versus the number of antennas $M$.}
        \label{NRMSEvsM}
    \end{subfigure}
\caption{NRMSE comparison of two schemes.}\label{fig:totalNRMSE}
\end{figure}

To compare the channel estimation performance of our proposed scheme with
the traditional scheme based on channel training in \cite{pilotestimation},
Fig. \ref{NRMSEvsPt} and Fig. \ref{NRMSEvsM} plot the normalized root mean-squared error (NRMSE) and their theoretical values for two methods versus ${\bar{P}}_{t}$ and the number of antennas, respectively.
NRMSE is defined as $\mathrm{NRMSE}=\sqrt{\frac{\mathrm{MMSE}}{\sum\limits_{s=1}^{S}{{{\left\| \mathbf{h} \right\|}^{2}}}}}$.
Since for the proposed channel estimation method enabled by ISSAC, the theoretical value of the MMSE of channel $\mathbf{h}$ is the same for both the digital and hybrid structure,
the theoretical value of $\mathrm{MMSE}$ takes from $e_{\mathrm{con}}$ in \eqref{RMSEe1} and $e_{\mathrm{eff}}$ in \eqref{elp}.
It is firstly observed from Fig. \ref{fig:totalNRMSE} that the actual values of both schemes are very close to the theoretical values.
In particular, it is demonstrated that for the conventional scheme, the channel error is independent of the number of antennas.
In contrast, it decreases as the number of antennas increases for our proposed scheme.
Further, compared with the conventional method, the NRMSE of the channel estimation scheme enabled by ISSAC is much smaller,
and the gap increases with the number of antennas, which makes our proposed scheme practically appealing for large-antenna systems.

\begin{figure}[!t]
  \centering
  \centerline{\includegraphics[width=3.1in,height=2.3in]{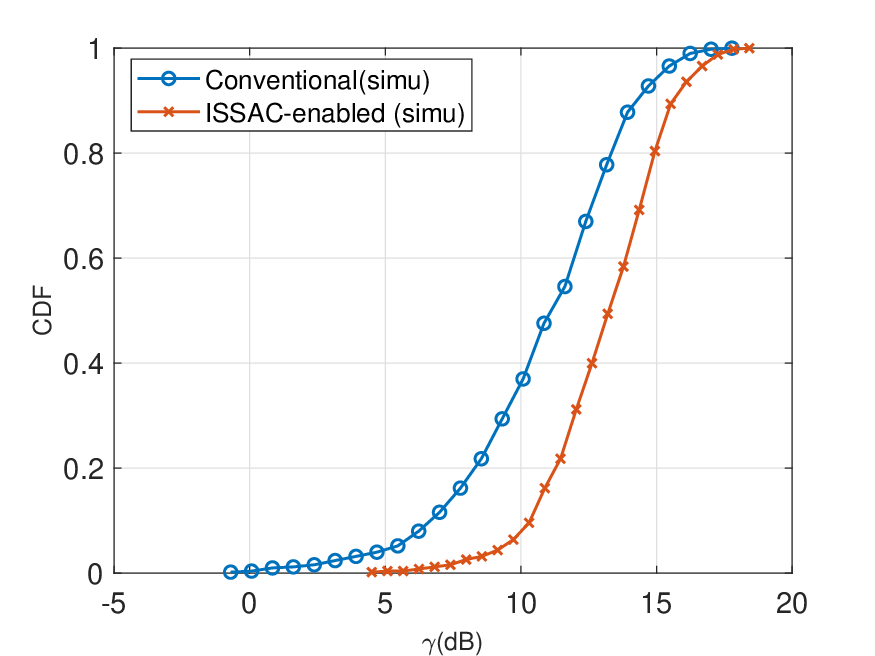}}
  \caption{CDF of receive SNR.}
  \label{CDF}
  \end{figure}

Fig. \ref{CDF} plots the cumulative distribution function (CDF) of the expected SNR of the signal for $\gamma_\mathrm{con}$ in \eqref{gamma1initial} and $\gamma_\mathrm{eff,hybrid}$ in \eqref{gammahybridorigin}.
It is observed that $\gamma_\mathrm{eff,hybrid}$ is larger than $\gamma_\mathrm{con}$ with the consideration of 90-percentile SNR.
It is revealed that even in the hybrid beamforming structure that uses limited RF chains,
the SNR performance of data transmission of channel estimation scheme enabled by ISSAC is much better than that of the traditional scheme in the digital beamforming structure using a large number of RF chains.

\begin{figure}[htbp]
  \centering
    \begin{subfigure}{0.45\textwidth}
      \centering
      \includegraphics[width=1\linewidth]{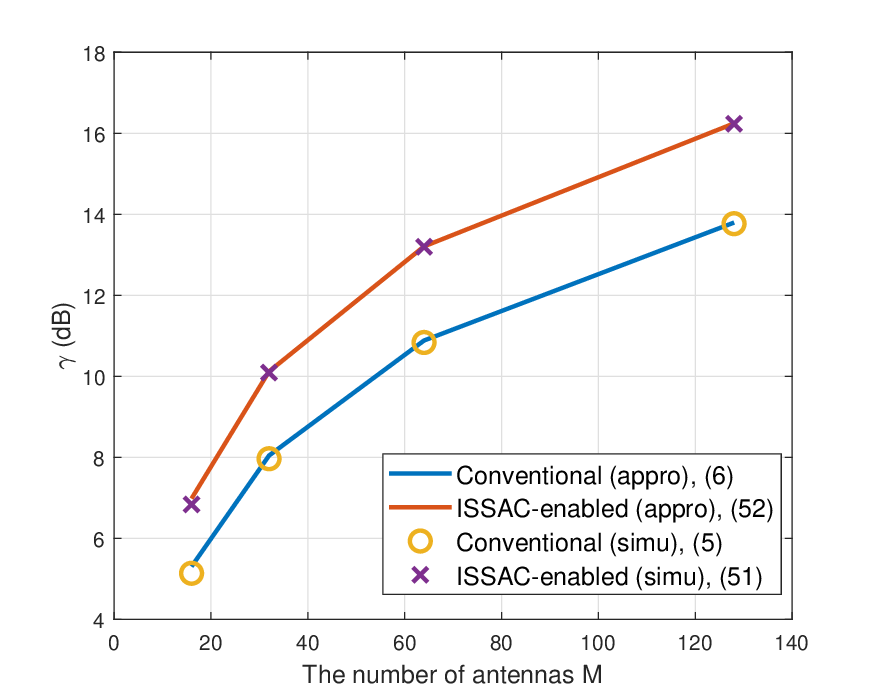}
        \caption{Receive SNR during data transmission phase versus the number of antennas $M$.}
        \label{hybridgammavsM}
    \end{subfigure}   
     \begin{subfigure}{0.45\textwidth}
      \centering
      \includegraphics[width=1\linewidth]{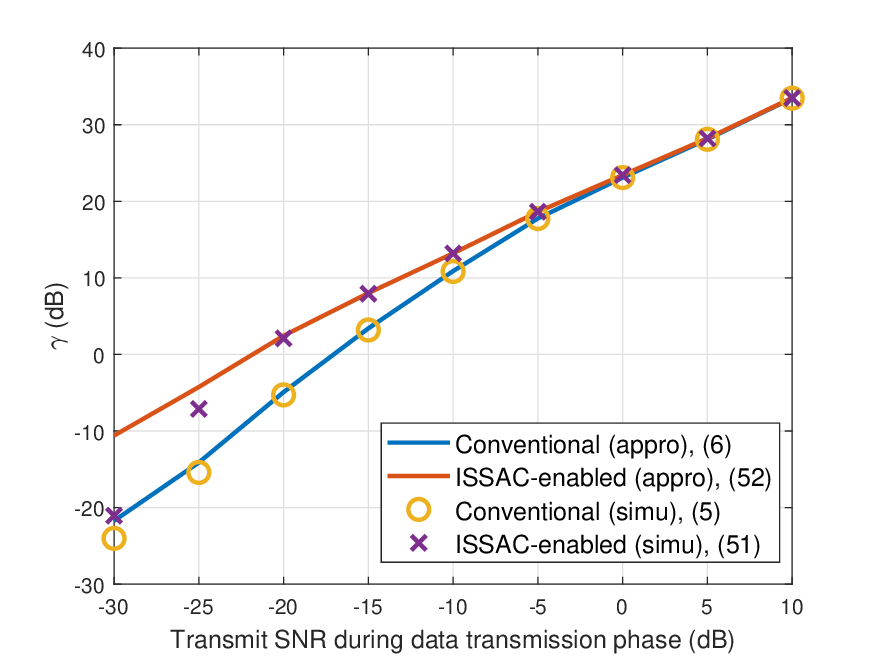}
        \caption{Receive SNR versus transmit SNR ${\bar{P}}_{d}$ during data transmission phase.}
        \label{hybridgammavsPd}
    \end{subfigure}
\caption{Receive SNR comparison of two schemes.}\label{fig:totalSNR}
\end{figure}

Fig. \ref{hybridgammavsM} and Fig. \ref{hybridgammavsPd} plot the receive SNR when decoding data signals for two schemes in \eqref{gamma1initial} and \eqref{gammahybridorigin} and their approximation values in \eqref{gamma1app} and \eqref{gammalphybrid} versus $M$ and ${\bar{P}}_{d}$, respectively.
It is firstly shown from Fig. \ref{hybridgammavsM} that the approximation values in \eqref{gamma1app} and \eqref{gammalphybrid} precisely match with the exact values in \eqref{gamma1initial} and \eqref{gammahybridorigin}.
It can also be observed in Fig. \ref{hybridgammavsPd} that when ${{\bar{P}}_{d}}$ is less than -5dB, the receive SNR for our proposed scheme is much higher than that of the traditional scheme, and this gap is largest at around -25dB.
This demonstrates that when SNR is low, our proposed scheme can provide more accurate channel estimation.

\begin{figure}[!t]
  \centering
  \centerline{\includegraphics[width=3.1in,height=2.3in]{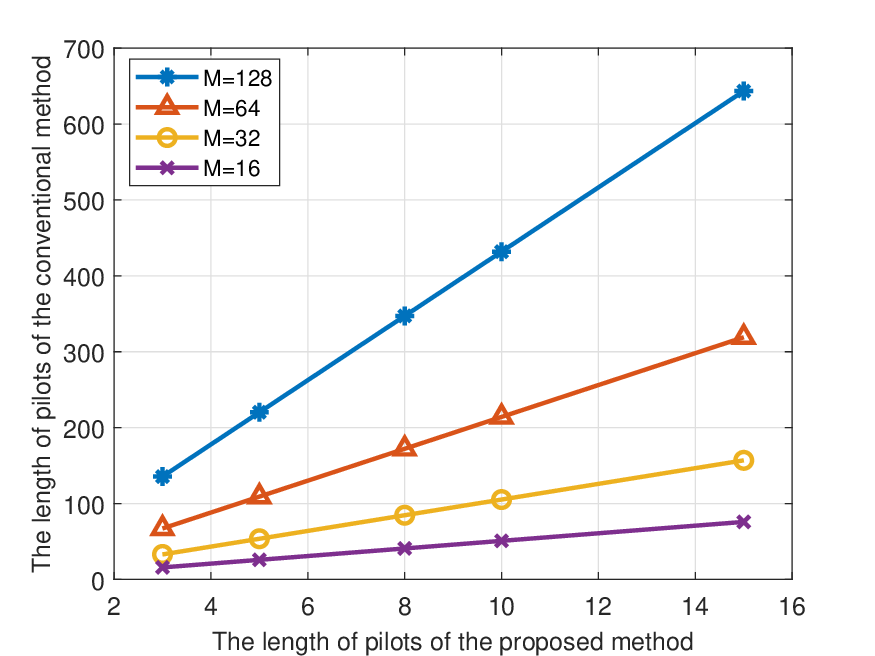}}
  \caption{The length of pilots of the conventional method versus that of the proposed method.}
  \label{rhocompare}
  \end{figure}

Fig. \ref{rhocompare} plots the pilot overhead of the conventional method ${{\rho }_{\mathrm{con}}}$ versus that of the proposed method ${{\rho }_{\mathrm{eff,hybrid}}}$ in achieving the same communication performance \eqref{pilotcompare}.
The numerical simulation is obtained by averaging the results of 1000 independent channels.
It can be firstly observed that the pilot overhead of the traditional scheme is in general higher by more than an order of magnitude than that of our proposed method.
In addition, it is observed that as the number of antennas $M$ increases,
the slope of the lines increases, which indicates that the advantage of our proposed scheme enabled by ISSAC in reducing pilot overhead is more significant in large-antenna systems.

\begin{figure}[!t]
  \centering
  \centerline{\includegraphics[width=3.1in,height=2.3in]{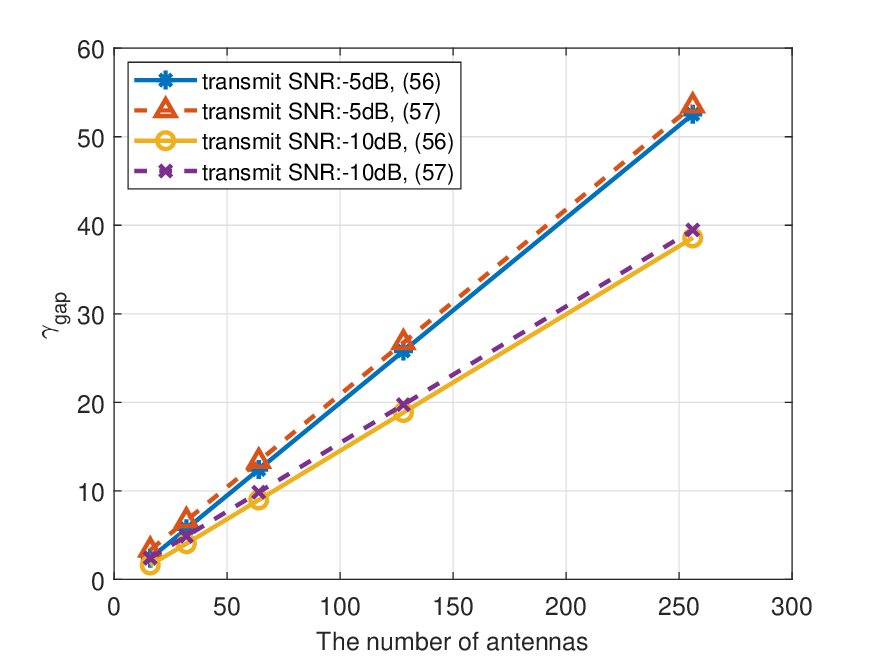}}
  \caption{The performance gap of the expected SNR versus the number of antennas.}
  \label{gapcompare}
  \end{figure}
Fig. \ref{gapcompare} plots the performance gap of the expected SNR between the proposed scheme and
the traditional scheme in \eqref{twomethodsgap} and its asymptotic value in \eqref{gapappro} versus the number of antennas $M$ for two transmit SNR values, i.e., ${{\bar{P}}_{t}}={{\bar{P}}_{d}}=-5$ dB and ${{\bar{P}}_{t}}={{\bar{P}}_{d}}=-10$ dB.
It can be firstly observed that the asymptotic values are close to the theoretical values.
Furthermore, it is also observed that the performance gap between two methods increases almost linearly with the number of antennas,
which indicates that our proposed method can achieve much better performance especially when the number of antennas is large.
\section{Conclusion}
In this paper, we propose an efficient channel estimation scheme enabled by ISSAC in the hybrid analog/digital beamforming structure, which requires very few pilots under the guarantee of accurate CSI estimation.
In particular, we first estimate the angles of the multi-path channel components by super-resolution algorithms with no dedicated pilot while communication data symbols are sent.
Then, with the obtained angles, by utilizing a few number of pilots, the multi-path channel coefficients are further estimated with the enjoyment of beamforming gain.
Both the LoS channel and more general multipath channels are studied
to evaluate the MMSE of channel estimation and the resulting beamforming gains.
It is revealed that the proposed scheme achieves more accurate channel estimation and higher receive SNR than the traditional pilot training method.
Simulation results are presented to validate our theoretical analysis.
\begin{appendices}
\section{Proof of Theorem 1}
It is non-trivial to derive the closed-form expression of \eqref{gamma1initial}.
In order to obtain some useful insights, $\mathbb{E}\left[ {{\mathbf{v}}_{\mathrm{con}}}\mathbf{v}_{\mathrm{con}}^{\mathrm{H}} \right]$ in \eqref{gamma1initial} can be approximated as
\begin{equation}
\begin{split}
  & \mathbb{E}\left[ {{\mathbf{v}}_{\mathrm{con}}}\mathbf{v}_{\mathrm{con}}^{\mathrm{H}} \right] =\mathbb{E}\bigg[ \frac{{{{\mathbf{\hat{h}}}}_{\mathrm{con}}}\mathbf{\hat{h}}_{\mathrm{con}}^{\mathrm{H}}}{{{\big\| {{{\mathbf{\hat{h}}}}_{\mathrm{con}}} \big\|}^{2}}} \bigg]\approx \frac{\mathbb{E}\left[ {{{\mathbf{\hat{h}}}}_{\mathrm{con}}}\mathbf{\hat{h}}_{\mathrm{con}}^{\mathrm{H}} \right]}{\mathbb{E}\big[ {{\big\| {{{\mathbf{\hat{h}}}}_{\mathrm{con}}} \big\|}^{2}} \big]} \\
 & =\frac{\mathbb{E}\left[ \left( \mathbf{h}+{{{\mathbf{\tilde{h}}}}_{\mathrm{con}}} \right){{\left( \mathbf{h}+{{{\mathbf{\tilde{h}}}}_{\mathrm{con}}} \right)}^{\mathrm{H}}} \right]}{\mathbb{E}\left[ {{\left\| \mathbf{h}+{{{\mathbf{\tilde{h}}}}_{\mathrm{con}}} \right\|}^{2}} \right]} =\frac{\mathbf{h}{{\mathbf{h}}^{\mathrm{H}}}+\frac{{{\sigma }^{2}}}{{{P}_{t}}\rho }{{\mathbf{I}}_{M}}}{{{\left\| \mathbf{h} \right\|}^{2}}+\frac{M{{\sigma }^{2}}}{{{P}_{t}}\rho }},\label{EvvH1}
\end{split}
\end{equation}
where the approximation holds due to $M\gg 1$.
Based on \eqref{EvvH1}, the expected SNR of the signal in \eqref{gamma1initial} is further approximated as
\begin{equation}
\begin{split}
  {{\gamma }_{\mathrm{con}}}& \approx\frac{{{P}_{d}}{{\left\| \mathbf{h} \right\|}^{2}}\left( \frac{{{P}_{t}}\rho }{M{{\sigma }^{2}}}{{\left\| \mathbf{h} \right\|}^{2}}+\frac{1}{M} \right)}{{{\sigma }^{2}}\left( \frac{{{P}_{t}}\rho }{M{{\sigma }^{2}}}{{\left\| \mathbf{h} \right\|}^{2}}+1 \right)} \\
 & =\frac{{{P}_{d}}{{\left\| \mathbf{h} \right\|}^{2}}\left( \rho \mathrm{SNR}_{t}+\frac{1}{M} \right)}{{{\sigma }^{2}}\left( \rho \mathrm{SNR}_{t}+1 \right)}\\
 & =\frac{{{P}_{d}}{{\left\| \mathbf{h} \right\|}^{2}}}{{{\sigma }^{2}}}\left( 1-\frac{1-\frac{1}{M}}{\rho \mathrm{SNR}_{t}+1} \right).
\end{split}
\end{equation}
The proof of Theorem 1 is completed.

\section{Proof of Theorem 3}
The MMSE of $\mathbf{h}$ for our proposed scheme can be written as
\begin{equation}
\begin{split}
  {{e}_{\mathrm{eff}}}& =\mathbb{E}\left[ {{\left\| {{{\mathbf{\tilde{h}}}}_{\mathrm{eff}}} \right\|}^{2}} \right] \\
 & =\mathbb{E}\bigg[ {{\bigg\| \frac{1}{\sqrt{{{P}_{t}}{{\rho }^{2}}}}\mathbf{A}\left( {\bm{\hat{\Theta }}} \right){{\left( {{\mathbf{A}}^{\mathrm{H}}}\left( {\bm{\hat{\Theta }}} \right)\mathbf{A}\left( {\bm{\hat{\Theta }}} \right) \right)}^{-1}}{{\mathbf{u}}_{t,\mathrm{mp}}} \bigg\|}^{2}} \bigg] \\
 & =\frac{1}{{{P}_{t}}{{\rho }^{2}}}\mathbb{E}\left[ \mathbf{u}_{t,\mathrm{mp}}^{\mathrm{H}}{{\left( {{\mathbf{A}}^{\mathrm{H}}}\left( {\bm{\hat{\Theta }}} \right)\mathbf{A}\left( {\bm{\hat{\Theta }}} \right) \right)}^{-1}}{{\mathbf{u}}_{t,\mathrm{mp}}} \right] \\
 & =\frac{1}{{{P}_{t}}{{\rho }^{2}}}\mathbb{E}\left[ \mathrm{tr}\left( \mathbf{u}_{t,\mathrm{mp}}^{\mathrm{H}}{{\left( {{\mathbf{A}}^{\mathrm{H}}}\left( {\bm{\hat{\Theta }}} \right)\mathbf{A}\left( {\bm{\hat{\Theta }}} \right) \right)}^{-1}}{{\mathbf{u}}_{t,\mathrm{mp}}} \right) \right] \\
 & =\frac{1}{{{P}_{t}}{{\rho }^{2}}}\mathbb{E}\left[ \mathrm{tr}\left( {{\left( {{\mathbf{A}}^{\mathrm{H}}}\left( {\bm{\hat{\Theta }}} \right)\mathbf{A}\left( {\bm{\hat{\Theta }}} \right) \right)}^{-1}}{{\mathbf{u}}_{t,\mathrm{mp}}}\mathbf{u}_{t,\mathrm{mp}}^{\mathrm{H}} \right) \right] \\
 & =\frac{1}{{{P}_{t}}{{\rho }^{2}}}\mathrm{tr}\Big( {{\left( {{\mathbf{A}}^{\mathrm{H}}}\left( {\bm{\hat{\Theta }}} \right)\mathbf{A}\left( {\bm{\hat{\Theta }}} \right) \right)}^{-1}}\mathbb{E}\left[ {{\mathbf{u}}_{t,\mathrm{mp}}}\mathbf{u}_{t,\mathrm{mp}}^{\mathrm{H}} \right] \Big),
\end{split}
\end{equation}
in which the second last equality holds for the identity $\mathrm{tr}(\mathbf{AB})=\mathrm{tr}(\mathbf{BA})$.
According to \cite{zhengjiaoboshu}, if the azimuth AoAs in ULA system are independently generated from a continuous distribution,
as the number of antenna $M$ approaches infinity and that of the channel paths $L$ satisfies $L=o\left( M \right)$,
the receive array response vectors tends to be orthogonal, i.e.,
$\mathbf{a}\left( {{\theta }_{l}} \right)\bot \mathrm{span}\left( \left\{ \mathbf{a}\left( {{\theta }_{k}} \right)|\forall k\ne l \right\} \right)$.

Therefore, ${{\mathbf{A}}^{\mathrm{H}}}\big( {\mathbf{\hat{\Theta }}} \big)\mathbf{A}\big( {\mathbf{\hat{\Theta }}} \big)\to M{{\mathbf{I}}_{L}}$
for the antennas of BS tends to infinity.
Then ${{e}_{\mathrm{eff}}}$ can be expressed as
\begin{equation}
{{{e}_{\mathrm{eff}}}=\frac{M{{\sigma }^{2}}}{{{P}_{t}}\rho }\mathrm{tr}\bigg( \frac{1}{M}{{\mathbf{I}}_{L}} \bigg)=\frac{L{{\sigma }^{2}}}{{{P}_{t}}\rho }.}
\end{equation}
The proof of Theorem 3 is completed.

\section{Proof of Theorem 4}
It is non-trivial to derive the closed-form expression of \eqref{gammaulpmulti}.
Therefore, to gain some insights, we assume the angle is perfectly estimated and approximate $\mathbb{E}[{ {{\mathbf{v}}_{\mathrm{eff}}}\mathbf{v}_{\mathrm{eff}}^{\mathrm{H}}}]$
in \eqref{gammaulpmulti} as
\begin{equation}
\begin{split}
  & \mathbb{E}\left[ {{\mathbf{v}}_{\mathrm{eff}}}\mathbf{v}_{\mathrm{eff}}^{\mathrm{H}} \right]=\mathbb{E}\bigg[ \frac{\left( {{{\mathbf{\tilde{h}}}}_{\mathrm{eff}}}+\mathbf{h} \right){{\left( {{{\mathbf{\tilde{h}}}}_{\mathrm{eff}}}+\mathbf{h} \right)}^{\mathrm{H}}}}{{{\left\| {{{\mathbf{\tilde{h}}}}_{\mathrm{eff}}}+\mathbf{h} \right\|}^{2}}} \bigg] \\
 & \approx \frac{\mathbb{E}\left[ \left( {{{\mathbf{\tilde{h}}}}_{\mathrm{eff}}}+\mathbf{h} \right){{\left( {{{\mathbf{\tilde{h}}}}_{\mathrm{eff}}}+\mathbf{h} \right)}^{\mathrm{H}}} \right]}{\mathbb{E}\left[ {{\left\| {{{\mathbf{\tilde{h}}}}_{\mathrm{eff}}}+\mathbf{h} \right\|}^{2}} \right]} \\
 & =\frac{\mathbb{E}\left[ {{{\mathbf{\tilde{h}}}}_{\mathrm{eff}}}\mathbf{\tilde{h}}_{\mathrm{eff}}^{\mathrm{H}} \right]+\mathbf{h}{{\mathbf{h}}^{\mathrm{H}}}}{\mathbb{E}\left[ {{\left\| {{{\mathbf{\tilde{h}}}}_{\mathrm{eff}}} \right\|}^{2}} \right]+{{\left\| \mathbf{h} \right\|}^{2}}} \\
 & =\frac{\frac{{{\sigma }^{2}}}{{{P}_{t}}}\mathbf{A}\left( {\mathbf{\hat{\Theta }}} \right){{\left( {{\mathbf{A}}^{\mathrm{H}}}\left( {\mathbf{\hat{\Theta }}} \right)\mathbf{A}\left( {\mathbf{\hat{\Theta }}} \right) \right)}^{-1}}{{\mathbf{A}}^{\mathrm{H}}}\left( {\mathbf{\hat{\Theta }}} \right)+\mathbf{h}{{\mathbf{h}}^{\mathrm{H}}}}{\frac{L{{\sigma }^{2}}}{{{P}_{t}}}+{{\left\| \mathbf{h} \right\|}^{2}}}.\label{EvmultivH}
\end{split}
\end{equation}
Based on \eqref{EvmultivH}, the expected SNR of the
signal in \eqref{gammaulpmulti} is approximated as
\begin{footnotesize}
\begin{equation}
\begin{split}
  & {{\gamma }_{\mathrm{eff}}}=\frac{{{P}_{d}}}{{{\sigma }^{2}}}\frac{\frac{{{\sigma }^{2}}}{{{P}_{t}}}{{\mathbf{h}}^{\mathrm{H}}}\mathbf{A}\left( {\mathbf{\hat{\Theta }}} \right){{\left( {{\mathbf{A}}^{\mathrm{H}}}\left( {\mathbf{\hat{\Theta }}} \right)\mathbf{A}\left( {\mathbf{\hat{\Theta }}} \right) \right)}^{-1}}{{\mathbf{A}}^{\mathrm{H}}}\left( {\mathbf{\hat{\Theta }}} \right)\mathbf{h}+{{\left\| \mathbf{h} \right\|}^{4}}}{\frac{L{{\sigma }^{2}}}{{{P}_{t}}}+{{\left\| \mathbf{h} \right\|}^{2}}} \\
 & =\frac{{{P}_{d}}}{{{\sigma }^{2}}}\frac{\frac{{{\sigma }^{2}}}{{{P}_{t}}}\mathrm{tr}\left( \mathbf{A}\left( {\mathbf{\hat{\Theta }}} \right){{\left( {{\mathbf{A}}^{\mathrm{H}}}\left( {\mathbf{\hat{\Theta }}} \right)\mathbf{A}\left( {\mathbf{\hat{\Theta }}} \right) \right)}^{-1}}{{\mathbf{A}}^{\mathrm{H}}}\left( {\mathbf{\hat{\Theta }}} \right)\mathbf{h}{{\mathbf{h}}^{\mathrm{H}}} \right)+{{\left\| \mathbf{h} \right\|}^{4}}}{\frac{L{{\sigma }^{2}}}{{{P}_{t}}}+{{\left\| \mathbf{h} \right\|}^{2}}} \\
 & \le \frac{{{P}_{d}}}{{{\sigma }^{2}}}\frac{\frac{{{\sigma }^{2}}}{{{P}_{t}}}\mathrm{tr}\left( \mathbf{A}\left( {\mathbf{\hat{\Theta }}} \right){{\left( {{\mathbf{A}}^{\mathrm{H}}}\left( {\mathbf{\hat{\Theta }}} \right)\mathbf{A}\left( {\mathbf{\hat{\Theta }}} \right) \right)}^{-1}}{{\mathbf{A}}^{\mathrm{H}}}\left( {\mathbf{\hat{\Theta }}} \right) \right)\mathrm{tr}\left( \mathbf{h}{{\mathbf{h}}^{\mathrm{H}}} \right)+{{\left\| \mathbf{h} \right\|}^{4}}}{\frac{L{{\sigma }^{2}}}{{{P}_{t}}}+{{\left\| \mathbf{h} \right\|}^{2}}} \\
 & =\frac{{{P}_{d}}}{{{\sigma }^{2}}}\frac{\frac{L{{\sigma }^{2}}}{{{P}_{t}}}{{\left\| \mathbf{h} \right\|}^{2}}+{{\left\| \mathbf{h} \right\|}^{4}}}{\frac{L{{\sigma }^{2}}}{{{P}_{t}}}+{{\left\| \mathbf{h} \right\|}^{2}}} \\
 & =\frac{{{P}_{d}}}{{{\sigma }^{2}}}{{\left\| \mathbf{h} \right\|}^{2}}.
\end{split}
\end{equation}
\end{footnotesize}
Then the upper bound of the expected SNR is obtained as
\begin{equation}
  {{{\gamma }_{\mathrm{eff}}} \le {{\gamma }_{\mathrm{eff,upper}}}=\frac{{{P}_{d}}}{{{\sigma }^{2}}}{{\left\| \mathbf{h} \right\|}^{2}}}.
\end{equation}
The proof of Theorem 4 is completed.

\section{Proof of Theorem 5}
It is non-trivial to derive the closed-form expression in \eqref{gammalphybrid}.
In order to gain some insights, we assume that the angle is perfectly
estimated, according to \eqref{bmalpha}, the estimation error of $\bm{\hat{\alpha }}$ is written as
\begin{equation}
\begin{split}
  \bm{\tilde{\alpha }}& =\bm{\hat{\alpha }}-\bm{\alpha } \\
 & =\frac{\sqrt{M}}{\sqrt{{{P}_{t}}}}{{({{\mathbf{A}}^{\mathrm{H}}}(\bm{\hat{\Theta }})\mathbf{A}(\mathbf{\hat{\Theta }}))}^{-1}}{{\mathbf{y}}_{t,\mathrm{mp}}}-\bm{\alpha } \\
 & =\frac{1}{\sqrt{{{P}_{t}}{{\rho }^{2}}}}{{({{\mathbf{A}}^{\mathrm{H}}}(\mathbf{\hat{\Theta }})\mathbf{A}(\mathbf{\hat{\Theta }}))}^{-1}}{{\mathbf{A}}^{\mathrm{H}}}(\mathbf{\hat{\Theta }}){{\mathbf{u}}_{t,\mathrm{mp}}} \\
 & \approx \frac{1}{M\sqrt{{{P}_{t}}{{\rho }^{2}}}}{{\mathbf{A}}^{\mathrm{H}}}(\mathbf{\hat{\Theta }}){{\mathbf{u}}_{t,\mathrm{mp}}}.
\end{split}
\end{equation}
The expected SNR of the signal in \eqref{yhybrid} is rewritten as
\begin{equation}
\begin{split}
  {{\gamma }_{\mathrm{eff,hybrid}}}& =\mathbb{E}\left[ \frac{{{\left| \sqrt{{{P}_{d}}M}{{{\bm{\hat{\alpha }}}}^{\mathrm{H}}}\bm{\alpha } \right|}^{2}}}{{{\sigma }^{2}}{{\left\| {\bm{\hat{\alpha }}} \right\|}^{2}}} \right]\\
 & \approx \frac{{{P}_{d}}M}{{{\sigma }^{2}}}\frac{\mathbb{E}\left[ {{\left| {{\left( \bm{\alpha }+\bm{\tilde{\alpha }} \right)}^{\mathrm{H}}}\bm{\alpha } \right|}^{2}} \right]}{\mathbb{E}\left[ {{\left\| \bm{\alpha }+\bm{\tilde{\alpha }} \right\|}^{2}} \right]},\label{SNRhybrid}
\end{split}
\end{equation}
where
\begin{equation}
\begin{split}
  & \mathbb{E}\left[ {{\left| {{\left( \bm{\alpha }+\bm{\tilde{\alpha }} \right)}^{\mathrm{H}}}\bm{\alpha } \right|}^{2}} \right] \\
 & =\mathbb{E}\bigg[ {{\bigg| {{\left\| \bm{\alpha } \right\|}^{2}}+\frac{1}{M\sqrt{{{P}_{t}}{{\rho }^{2}}}}\mathbf{{u}}_{t,\mathrm{mp}}^{\mathrm{H}}\mathbf{A}\left( {\bm{\hat{\Theta }}} \right)\bm{\alpha } \bigg|}^{2}} \bigg] \\
 & ={{\left\| \bm{\alpha } \right\|}^{4}}+\frac{{{\sigma }^{2}}{{\left\| \bm{\alpha } \right\|}^{2}}}{M{{P}_{t}}\rho },
\end{split}
\end{equation}
and
\begin{equation}
\begin{split}
  & \mathbb{E}\left[ {{\left\| \bm{\alpha }+\bm{\tilde{\alpha }} \right\|}^{2}} \right] \\
 & ={{\left\| \bm{\alpha } \right\|}^{2}}+\mathbb{E}\left[ \frac{1}{{{M}^{2}}{{P}_{t}}{{\rho }^{2}}}\mathbf{u}_{t,\mathrm{mp}}^{\mathrm{H}}\mathbf{A}\left( {\bm{\hat{\Theta }}} \right){{\mathbf{A}}^{\mathrm{H}}}\left( {\bm{\hat{\Theta }}} \right){{\mathbf{u}}_{t,\mathrm{mp}}} \right] \\
 & ={{\left\| \bm{\alpha } \right\|}^{2}}+\frac{1}{{{M}^{2}}{{P}_{t}}{{\rho }^{2}}}\mathbb{E}\left[ \mathrm{tr}\left( \mathbf{{u}}_{t,\mathrm{mp}}^{\mathrm{H}}\mathbf{A}\left( {\mathbf{\hat{\Theta }}} \right){{\mathbf{A}}^{\mathrm{H}}}\left( {\mathbf{\hat{\Theta }}} \right){{\mathbf{u}}_{t,\mathrm{mp}}} \right) \right] \\
 & ={{\left\| \bm{\alpha } \right\|}^{2}}+\frac{1}{{{M}^{2}}{{P}_{t}}{{\rho }^{2}}}\mathbb{E}\left[ \mathrm{tr}\left( {{\mathbf{A}}^{\mathrm{H}}}\left( {\mathbf{\hat{\Theta }}} \right){{\mathbf{u}}_{t,\mathrm{mp}}}\mathbf{{u}}_{t,\mathrm{mp}}^{\mathrm{H}}\mathbf{A}\left( {\mathbf{\hat{\Theta }}} \right) \right) \right] \\
 & ={{\left\| \bm{\alpha } \right\|}^{2}}+\frac{1}{{{M}^{2}}{{P}_{t}}{{\rho }^{2}}}\mathrm{tr}\left( {{\mathbf{A}}^{\mathrm{H}}}\left( {\mathbf{\hat{\Theta }}} \right)\mathbb{E}\left[ {{\mathbf{u}}_{t,\mathrm{mp}}}\mathbf{{u}}_{t,\mathrm{mp}}^{\mathrm{H}} \right]\mathbf{A}\left( {\mathbf{\hat{\Theta }}} \right) \right) \\
 & ={{\left\| \bm{\alpha } \right\|}^{2}}+\frac{{{\sigma }^{2}}}{{{M}^{2}}{{P}_{t}}\rho }\mathrm{tr}\left( {{\mathbf{A}}^{\mathrm{H}}}\left( {\mathbf{\hat{\Theta }}} \right)\mathbf{A}\left( {\mathbf{\hat{\Theta }}} \right) \right) \\
 & ={{\left\| \bm{\alpha } \right\|}^{2}}+\frac{{{\sigma }^{2}}L}{M{{P}_{t}}\rho }.
\end{split}
\end{equation}
Then \eqref{SNRhybrid} can be rewritten as
\begin{equation}
{{{\gamma }_{\mathrm{eff,hybrid}}}\approx \frac{{{P}_{d}}M{{\left\| \bm{\alpha } \right\|}^{2}}}{{{\sigma }^{2}}}\left( \frac{{{\left\| \bm{\alpha } \right\|}^{2}}+\frac{{{\sigma }^{2}}}{M{{P}_{t}}\rho }}{{{\left\| \bm{\alpha } \right\|}^{2}}+\frac{{{\sigma }^{2}}L}{M{{P}_{t}}\rho }} \right)}.
\end{equation}
The proof of Theorem 5 is completed.
\end{appendices}

\end{document}